\documentclass[twocolumn,superscriptaddress, aps, prd]{revtex4-1}
\usepackage{gnuplottex}
\usepackage{mathtools}
\usepackage{float}
\usepackage{array}
\usepackage[english]{babel}
\usepackage[utf8]{inputenc}
\usepackage{amsmath}
\usepackage[table]{xcolor}
\usepackage{amsfonts}
\usepackage{graphicx}
\usepackage{dsfont}
\usepackage[%
  colorlinks=true,
  urlcolor=blue,
  linkcolor=blue,
  citecolor=blue
]{hyperref}
\usepackage{algorithm}
\usepackage[noend]{algpseudocode}
\usepackage{bbold}
\usepackage{mathtools}
\usepackage{amssymb}
\usepackage{bm}
\usepackage{bbm}
\usepackage{tabularx, booktabs}
\usepackage{xspace}
\usepackage{listings}
\usepackage{lipsum}
\usepackage{chemformula}
\usepackage{tikz}
\usepackage{physics}
\usepackage{mathtools}
\usepackage{siunitx}

\newcolumntype{Y}{>{\centering\arraybackslash}X}
\definecolor{dgreen}{rgb}{0,.5,0}
\definecolor{dblue}{rgb}{0,0,.5}
\definecolor{dred}{rgb}{0.5,0,.5}

\usepackage{soul}

\usepackage{scalerel,stackengine}
\stackMath
\newcommand\reallywidehat[1]{%
\savestack{\tmpbox}{\stretchto{%
  \scaleto{%
    \scalerel*[\widthof{\ensuremath{#1}}]{\kern-.6pt\bigwedge\kern-.6pt}%
    {\rule[-\textheight/2]{1ex}{\textheight}}%WIDTH-LIMITED BIG WEDGE
  }{\textheight}% 
}{0.5ex}}%
\stackon[1pt]{#1}{\tmpbox}%
}
\newcommand{\bfr}{\mathbf{r}}
\newcommand{\bfR}{\mathbf{R}}

\def\ddroit{{\rm d}}
\newcommand*{\thead}[1]{\multicolumn{1}{|c|}{\bfseries #1}}

\AtBeginDocument{%
    \newwrite\bibnotes
    \def\bibnotesext{Notes.bib}
    \immediate\openout\bibnotes=\jobname\bibnotesext
    \immediate\write\bibnotes{@CONTROL{REVTEX41Control}}
    \immediate\write\bibnotes{@CONTROL{%
    apsrev41Control,author="08",editor="1",pages="1",title="0",year="1"}}
     \if@filesw
     \immediate\write\@auxout{\string\citation{apsrev41Control}}%
    \fi
}%

\begin{document}

% \title{Quantum chemistry strategies to reduce the 1-norm of the electronic structure Hamiltonian for quantum computing applications
% } % Article title

\title{ Orbital transformations to reduce the 1-norm of the electronic structure Hamiltonian for quantum computing applications } % Article title

\author{Emiel Koridon} 
\affiliation{Instituut-Lorentz, Universiteit Leiden, P.O. Box 9506, 2300 RA Leiden, The Netherlands}
\affiliation{Theoretical Chemistry, Vrije Universiteit, De Boelelaan 1083, NL-1081 HV, Amsterdam, The Netherlands}
\author{Saad Yalouz}
\email{yalouzsaad@gmail.com}
\affiliation{Theoretical Chemistry, Vrije Universiteit, De Boelelaan 1083, NL-1081 HV, Amsterdam, The Netherlands}
\affiliation{Instituut-Lorentz, Universiteit Leiden, P.O. Box 9506, 2300 RA Leiden, The Netherlands}
\author{Bruno Senjean}
\email{bruno.senjean@umontpellier.fr}
\affiliation{ICGM, Univ Montpellier, CNRS, ENSCM, Montpellier, France}
\affiliation{Instituut-Lorentz, Universiteit Leiden, P.O. Box 9506, 2300 RA Leiden, The Netherlands}
\author{Francesco Buda}
\affiliation{Leiden Institute of Chemistry,
Leiden University, Einsteinweg 55, P.O. Box 9502, 2300 RA Leiden, The Netherlands.}
\author{Thomas E. O'Brien}
\affiliation{Google Research, 80636 Munich, Germany}
\affiliation{Instituut-Lorentz, Universiteit Leiden, P.O. Box 9506, 2300 RA Leiden, The Netherlands}
\author{Lucas Visscher} 
\affiliation{Theoretical Chemistry, Vrije Universiteit, De Boelelaan 1083, NL-1081 HV, Amsterdam, The Netherlands}

\begin{abstract}\def\ddroit{{\rm d}}

Reducing the complexity of quantum algorithms to treat quantum chemistry problems is essential to demonstrate an eventual quantum advantage of Noisy-Intermediate Scale Quantum (NISQ) devices over their classical counterpart.
Significant improvements have been made recently to simulate the time-evolution operator $U(t) = e^{i\mathcal{\hat{H}}t}$ where $\mathcal{\hat{H}}$ is the electronic structure Hamiltonian, or to simulate $\mathcal{\hat{H}}$ directly (when written as a linear combination of unitaries) by using block encoding or ``qubitization'' techniques.
A fundamental measure quantifying the practical implementation complexity of these quantum algorithms is the so-called ``1-norm'' of the qubit-representation of the Hamiltonian,
which can be reduced by writing the Hamiltonian in factorized or tensor-hypercontracted forms for instance.
In this work, we investigate the effect of classical pre-optimization of the electronic structure Hamiltonian representation, via single-particle basis transformation, on the 1-norm.
Specifically, we employ several localization schemes and benchmark the 1-norm of several systems of different sizes (number of atoms and active space sizes).
We also derive a new formula for the 1-norm as a function of the electronic integrals, and use this quantity as a cost function for an orbital-optimization scheme that improves over localization schemes.
This paper gives more insights about the importance of the 1-norm in quantum computing for quantum chemistry, and provides simple ways of decreasing its value to reduce the complexity of quantum algorithms.

\end{abstract}

\maketitle

\section{Introduction}

Quantum chemistry has been identified as the killer application of quantum computers, which promise to solve problems of high industrial impact that are not tractable for their classical counterparts~\cite{reiher2017elucidating,cao2019quantum,bauer2020quantum,von2020quantum}.
Unfortunately, quantum devices are noisy and only shallow circuits can be implemented with a relatively accurate fidelity.
Hence, even though we know how to solve the electronic structure problem with, for instance, the quantum phase estimation (QPE)~\cite{kitaev1995quantum,abrams1997simulation,abrams1999quantum,aspuru2005simulated,obrien2019quantum} algorithm, this remains out of reach in practice and one has to come up with new algorithms that can actually be run within this noisy intermediate-scale quantum (NISQ) era~\cite{Preskill_2018}.
Reducing circuit depth has been achieved by interfacing quantum and classical devices, thus leading to hybrid quantum-classical algorithms such as the variational quantum eigensolver (VQE)~\cite{peruzzo2014variational}.
Since then, variational algorithms have been successfully applied to the estimation of ground-state (see Refs.~\onlinecite{mcclean2016theory,cerezo2020variational,endo2021hybrid,bharti2021noisy} and references therein) and excited-state energies~\cite{mcclean2017hybrid,lee2018generalized,ollitrault2020quantum,nakanishi2019subspace,parrish2019hybrid,parrish2019quantum,ibe2020calculating,higgott2019variational,jones2019variational,jouzdani2019method,yalouz2021state}, as well as molecular properties~\cite{kassal2009quantum,obrien2019calculating,mitarai2020theory,sokolov2020microcanonical}.

However, it remains unclear if these algorithms can provide a clear quantum advantage in the long run, especially due to the difficulty in circuit optimization~\cite{bittel2021training} and to the large overhead in the number of measurements required to achieve sufficient accuracy~\cite{wecker2015progress,rubin2018application},
though significant progress has been made recently~\cite{kandala2017hardware,rubin2018application,izmaylov2019revising,verteletskyi2020measurement,jena2019pauli,izmaylov2019unitary,yen2020measuring,gokhale2019minimizing,huggins2021efficient,bonet2020nearly,zhao2020measurement,wang2021minimizing,garcia2021learning}.
This led to the development of new strategies for reducing the resources required to implement quantum algorithms (such as gate complexity), for both the fault-tolerant and NISQ era.
For simulating the time-evolution operator $U(t) = e^{i\mathcal{\hat{H}}t}$
where $\mathcal{\hat{H}}$ is the electronic Hamiltonian, Campbell has shown that using randomized compiler is better suited than the Trotter-Suzuki decomposition, the so called quantum stochastic drift protocol (qDRIFT)~\cite{campbell2019random}.
One can also directly simulate the Hamiltonian by using linear combination of unitaries, generalized by the block encoding or ``qubitization'' formalism~\cite{childs2012hamiltonian,berry2014exponential,low2019hamiltonian,berry2019qubitization}.
Most of these algorithms have a gate complexity that scales with respect to the parameter $\lambda_Q = \sum_i | h_i |$, where $\mathcal{\hat{H}} = \sum_i h_i \hat{P}_i$ is the qubit Hamiltonian and $\hat{P}_i$ are Pauli strings~\cite{jordanwigner}.
Lowering the value of $\lambda_Q$ -- usually referred to the 1-norm of the Hamiltonian -- has a significant impact on quantum algorithms, and techniques such as double-factorization~\cite{von2020quantum},
tensor hypercontraction~\cite{lee2020even} and $n$-representability constraints~\cite{rubin2018application} have proven successful in this respect.

In this work, we investigate how the value of the 1-norm and its scaling with respect to the number of orbitals can be reduced by single-particle basis rotations.
As a starting point we consider the use of localized orbitals that are commonly available in most quantum chemistry codes. 
Generating such orbitals presents a simple and effortless way of reducing gate complexity by classical pre-optimization of the Hamiltonian representation.

We show that the use of localized orbitals has a significant impact on the 1-norm of all systems studied, from simple one-dimensional hydrogen chains (also studied in Refs.~\onlinecite{rubin2018application,lee2020even}) to much more complex organic and inorganic molecules.
Additionally, we connect the expressions of the 1-norm before and after the fermion-to-qubit mapping.
This expression is used to define a new cost function for single-particle basis rotations, and is shown to reduce the value of $\lambda_Q$ even more than standard localization schemes.

The paper is organized as follows.
After introducing the electronic structure problem in Sec.~\ref{sec:ESP},
the impact of the 1-norm on quantum algorithms is detailed in Sec.~\ref{sec:onenorm_QC}.
Then, we review the localization schemes applied in this work in Sec.~\ref{sec:loc}
followed by our so-called 1-norm orbital-optimization method in Sec.~\ref{sec:1-norm-opt}.
Finally, preceded by the computational details in Sec.~\ref{sec:comp}, the scaling of the 1-norm when increasing the number of atoms is investigated in Sec.~\ref{sec:HandC} for hydrogen and alkane chains, followed by a benchmark on several systems in Sec.~\ref{sec:benchmark} and a study of the effect of increasing the active space size in Sec.~\ref{sec:lambda_scaling}.
Conclusions and perspectives are given in Sec.~\ref{sec:conclusion}.

\section{Theory}\label{sec:theory}

\subsection{Electronic structure Hamiltonian in second quantization}\label{sec:ESP}

% In a molecular system composed, the Born-Oppenheimer approximation is usually employ to uncouple the dynamics of both type of particles. Nuclei which correspond to heavy and slow particles are usually considered as frozen in space. Electrons, which are lighter and faster, are quantum by nature and can instantaneously rearrange around the frozen nuclei.

In quantum chemistry, the second quantization formalism is usually employed to describe the electronic properties of molecules.
Under the Born-Oppenheimer approximation, $N_e$ electrons can rearrange around $N_a$ nuclei by occupying a restricted set of $N$ spatial molecular orbitals (MO). 
The orthonormal MO basis $\lbrace \phi_p (\bfr) \rbrace$ is built from linear combination of atomic orbitals (LCAO) and is usually obtained from an inexpensive mean-field calculation, \textit{e.g.} with the Hartree--Fock (HF) method~\cite{helgaker2014molecular}.
In this basis, the molecular Hamiltonian can be expressed in a spin-free form that reads (in atomic units)
\begin{eqnarray}
\hat{\mathcal{H}}  =  \sum_{pq}^{N} h_{pq} \hat{E}_{pq} + \dfrac{1}{2} \sum_{pqrs}^{N} g_{pqrs}  \hat{e}_{pqrs},
\label{eq:el_Ham}
\end{eqnarray}
where $\hat{E}_{pq} = \sum_{\sigma} \hat{a}_{p\sigma}^\dagger \hat{a}_{q\sigma}$ and $ \hat{e}_{pqrs} = \sum_{\sigma,\tau} \hat{a}_{p\sigma}^\dagger \hat{a}_{r\tau}^\dagger \hat{a}_{s\tau}\hat{a}_{q\sigma} $ are the one- and two-body spin-free operators. 
In Eq.~(\ref{eq:el_Ham}), the coefficients
\begin{equation}
h_{pq}  = \int \phi_p^*(\bfr_1 ) \left(- \dfrac{1}{2} \grad_{\bfr_1}^2 - \sum_{A=1}^{N_a}\dfrac{Z_A}{\bfr_{1A}} \right) \phi_q(\bfr_1 ) \ddroit \bfr_1
\end{equation}
are the so-called one-electron integrals which encode, for each individual electron, the associated kinetic energy and Coulombic interaction with the $N_a$ nuclei of the molecule (with  $\bfr_{iA} = | \bfr_i - \bfR_A |$ and $Z_A$ the distance between electron $i$ and nuclei $A$, and the atomic number of atom $A$).
The coefficients
\begin{equation}\label{eq:eri}
g_{pqrs}  = (pq|rs) = \iint  
\phi_p^*(\bfr_1 ) \phi_r^*(\bfr_2 )
\dfrac{1}{\bfr_{12}}\phi_q(\bfr_1 ) \phi_s(\bfr_2 )  \ddroit \bfr_1 \ddroit \bfr_2 
\end{equation}
are the so-called two-electron integrals encoding the Coulombic repulsion between each pair of electrons (with $\bfr_{ij} = | \bfr_i - \bfr_j |$ the distance between electrons $i$ and $j$).
Solving the electronic structure problem consists in solving the time-independent Schr\"{o}dinger equation
\begin{equation}
\mathcal{\hat{H}} \ket{\Psi_\ell} = E_\ell \ket{\Psi_\ell},
\end{equation}
with $\ket{\Psi_\ell} $ an electronic eigenstate with corresponding energies $E_\ell$.
This is only possible in the case of small molecules and small basis sets, due to the exponential scaling of the computational cost with respect to the size of the system.
A commonly employed approach is the so-called active space approximation which divides the MO space into three parts: the core (frozen occupied orbitals), the active, and the virtual spaces (deleted unoccupied orbitals), such that only the electrons inside the active space are treated explicitly.
This approximation will be used throughout this work and is equivalent to finding the eigenstates of an effective Hamiltonian $\hat{{\mathcal{H}}}^{FC}$ also called the ``frozen-core Hamiltonian''
(see Appendix.~\ref{appendix:FrozenCore}).

The electronic Hamiltonian can be mapped onto an appropriate representation for quantum computers by doing a fermion-to-qubit transformation (such as Jordan--Wigner~\cite{jordanwigner}), resulting in a linear combination of Pauli strings $\hat{P}_j\in\{I,X,Y,Z\}^{\otimes N}$,
\begin{eqnarray}\label{eq:H_qub}
    \hat{\mathcal{H}}_{Q} = \sum_j^{\mathcal{S}} h_j \hat{P}_j.
\end{eqnarray}
Here, $\mathcal{S}$ denotes the sparsity of the Hamiltonian and generally scales as $\mathcal{O}(N^4)$, but sometimes $\mathcal{O}(N^2)$ for a sufficiently large system and a localized basis~\cite{mcclean2014exploiting}.

\subsection{The 1-Norm in quantum computing}\label{sec:onenorm_QC}

\begin{table}
\begin{tabular}{|c|c|}
 \hline
Method &  Toffoli/T complexity \\
 \hline
Database method~[\onlinecite{babbush2016exponentially}] & $\mathcal{O}(N^4\lambda / \epsilon)$ \\[0pt]
qDRIFT~[\onlinecite{campbell2019random}] & $\mathcal{O}(\lambda^2 / \epsilon^2)$\\[0pt]
Qubitization (sparse method)~[\onlinecite{berry2019qubitization}]  & $\mathcal{O}((N+\sqrt{\mathcal{S}})\lambda / \epsilon)$\\[0pt]
Qubitization (single factorization)~[\onlinecite{berry2019qubitization}]  & $\mathcal{O}(N^{3/2}\lambda_{\rm SF} / \epsilon)$\\[0pt]
%Randomized compiled phase estimation~[\onlinecite{kivlichan2019phase}] &
%$\tilde{\mathcal{O}}(\lambda_V^2 / \epsilon^2)$\\[0pt]
Qubitization (double factorization)~[\onlinecite{von2020quantum}] &
$\mathcal{O}(N\lambda_{\rm DF}\sqrt{\Xi}/\epsilon)$\\[0pt]
Tensor hypercontraction~[\onlinecite{lee2020even}] & $\tilde{\mathcal{O}}(N\lambda_\zeta / \epsilon)$\\
 \hline
VQE & \# of measurements \\\hline
State tomography~[\onlinecite{peruzzo2014variational,wecker2015progress,mcclean2016theory}] &  $M \approx \lambda^2 / \epsilon^2$ \\
Basis rotation grouping~[\onlinecite{huggins2021efficient}] & $M\approx \lambda_{\rm SF}^2/\epsilon^2$\\
 \hline
\end{tabular}
\caption{Lowest asymptotic scaling of different quantum algorithms involving the $\lambda$ parameters. $\lambda$ denotes the 1-norm of the qubit Hamiltonian in Eq.~(\ref{eq:lambda}),
$\lambda_{\rm SF}$ (denoted $\lambda_W$ in Ref.~\onlinecite{berry2019qubitization}) the one of the singly-factorized fermionic Hamiltonian,
$\lambda_{\rm DF}$ the one of the doubly-factorised Hamiltonian and 
$\lambda_\zeta$ the one of the non-orthogonal tensor hypercontracted Hamiltonian representation.
$\mathcal{S}$ is the sparsity of the electronic Hamiltonian and
$\Xi$ the average rank of the second tensor factorization~\cite{motta2018low,von2020quantum}. See text for further details.}
\label{tab:scaling_history}
\end{table}

The term `1-norm' in quantum computing refers to a norm induced on a Hilbert-space operator by its decomposition as a sum of simpler terms.
To calculate a 1-norm, we write an operator $\mathcal{\hat{H}}$ (e.g. a Hamiltonian) in the form
\begin{equation}
  \mathcal{\hat{H}}=\sum_jb_j\hat{B}_j,
\end{equation}
where the $\hat{B}_j$ are operators on $\mathbb{C}^{2^{2N}}$, and the $b_j$ are complex numbers.
Typically the operators $\hat{B}_j$ are chosen to be either unitary ($\hat{B}_j^\dag \hat{B}_j=1$) or unital ($\hat{B}_j^\dag \hat{B}_j<1$).
We define the 1-norm of $\mathcal{\hat{H}}$ (induced by this decomposition) to be the 1-norm on the vector $\vec{b}$ (with components $b_j$):
\begin{equation}
    \lambda_B = \lambda_B(\mathcal{\hat{H}}) = \sum_j|b_j|.
\end{equation}
(It is common to suppress the dependence on the Hamiltonian $\mathcal{\hat{H}}$ when $\lambda_B$ is unambiguous.)
For example, under the above decomposition into Pauli operators [Eq.~(\ref{eq:H_qub})], we have
\begin{align}\label{eq:1norm}
    \lambda_Q \equiv ||\vec{h}||_1 = \sum_j |h_j|
\end{align}
In general, $\lambda_B$ is a norm on the space spanned by the set $\{\hat{B}_j\}$ as long as this set is linearly independent, as the mapping $\mathcal{\hat{H}}\leftrightarrow \vec{b}$ is then bijective and linear.
The Pauli operators $\hat{P}_j$ used above make a natural choice for $\hat{B}_j$ as they span the set of $2^{2N}\times 2^{2N}$ matrices, and so induce a norm for any operators on the Hilbert space.
Pauli operators are also unitary, and may be easily implemented in a quantum circuit~\cite{whitfield11simulation}, making them a common choice for Hamiltonian decompositions.
However, unlike other operator size measures, 1-norms may be heavily optimized by a change of operator basis (whereas \textit{e.g.} the 2-norm or infinity-norm are both invariant under unitary rotation). As operator 1-norms play a role in determining the cost of many quantum computing algorithms, this makes them a key target of study in quantum algorithm research.

A versatile method for implementing an arbitrary (non-Hermitian) operator on a quantum device is the linear combination of unitaries (LCU) method~\cite{childs2012hamiltonian}, which involves a decomposition over a set $\{\hat{B}_j\}$ of strictly unitary operators.
As $\mathcal{\hat{H}}$ is typically not unitary, the original LCU technique requires postselection, but this forms the basis for unitary methods such as qubitization~\cite{low2019hamiltonian}.
These in turn underpin many proposals for quantum computing on a quantum computer~\cite{motta2018low,von2020quantum,lee2020even}, and the dependence on $\lambda_B$ is pulled directly through.
We give a brief description here of the key identity in LCU methods to show where the dependence on $\lambda_B$ appears.
LCU methods all require a control register with states $|j\rangle$ that encode the index to the operator $\hat{B}_j$.
(One need only encode those operators $\hat{B}_j$ that are relevant to the target $\mathcal{\hat{H}}$.)
Then, these methods encode the coefficients of $\mathcal{\hat{H}}$ on the register by preparing the state $\frac{1}{\sqrt{\lambda_B}}\sum_j\sqrt{b_j}|j\rangle$.
Then, consider implementing the joint unitary $\sum_j|j\rangle\langle j|\otimes \hat{B}_j$ (that is, applying the unitary operator $\hat{B}_j$ to a system conditional on the control register being in the state $j$).
Post-selecting the control register returning to the initial state performs the LCU operation to the register
\begin{align}
    \frac{1}{\sqrt{\lambda_B}}\sum_j \sqrt{b_j}\langle j|&\left[\sum_j|j\rangle\langle j|\hat{B}_j\right]\frac{1}{\sqrt{\lambda_B}}\sum_j \sqrt{b_j}|j\rangle\nonumber\\
    &=\frac{1}{\lambda_B}\sum_jb_j\hat{B}_j=\frac{1}{\lambda_B}\mathcal{\hat{H}}.
\end{align}
We see here that $\lambda_B$ emerges naturally as the normalization constant of the LCU unitary.
This dependence is passed on to any LCU-based method, suggesting that minimizing $\lambda_B$ is key to optimizing any such techniques.

The 1-norm can also play a role in methods where Hamiltonian simulation is implemented stochastically.
For instance, in the qDRIFT~\cite{campbell2019random} method, hermitian terms $\hat{B}_j$ in the Hamiltonian are chosen at random, weighted by $|b_j|/\lambda_B$, and exponentiated on a system as $e^{i\tau \hat{B}_j}$
($\lambda_B$ appears here immediately as the normalization of the probability distribution).
To first order, this implements the channel on a state $\hat{\rho}$~\cite{campbell2019random}
\begin{equation}
    \mathcal{E}(\hat{\rho})= \hat{\rho} + i\sum_j\frac{b_j\tau}{\lambda_B}(\hat{B}_j\hat{\rho} - \hat{\rho} \hat{B}_j) + O(\tau^2),
\end{equation}
which can be observed to be the unitary evolution of $\hat{\rho}$ under $e^{i\mathcal{\hat{H}}dt}$ for $dt=\tau/\lambda_B$.
To approximate time evolution for time $t$ to error $\epsilon$ then requires repeating this process $O(2\lambda_B^2t^2/\epsilon)$ times.
This has been extended upon recently~\cite{ouyang2020compilation}, yielding a more complicated $\lambda_B$ dependence.

A final situation where the 1-norm of an operator plays a role is in tomography.
One may in principle measure the expectation value of an arbitrary Hermitian operator $\mathcal{\hat{H}}$ on a state $\rho$ by repeatedly preparing $\hat{\rho}$, rotating into the eigenbasis of $\mathcal{\hat{H}}$ and reading out the device.
The variance in averaging $M$ shots of preparation and measurement is directly given by the variance of $\mathcal{\hat{H}}$ on $\hat{\rho}$:
\begin{equation}
    \mathrm{Var}\left[\langle\mathcal{\hat{H}}\rangle\right] = \frac{1}{M}\left[\langle\mathcal{\hat{H}}^2\rangle - \langle\mathcal{\hat{H}}\rangle ^2\right].
\end{equation}
However, in practice (especially in the NISQ era) arbitrary rotations into an unknown basis may be costly.
As expectation values are linear, given a decomposition of $\mathcal{\hat{H}}$ over $\{\hat{B}_j\}$, one may instead perform $M_j$ different preparations of $\hat{\rho}$ and measurements in the basis of each $\hat{B}_j$ and sum the result,
\begin{equation}
    \langle\mathcal{\hat{H}}\rangle = \sum_j b_j\langle \hat{B}_j\rangle,
\end{equation}
with variance
\begin{equation}
    \mathrm{Var}\left[\langle\mathcal{\hat{H}}\rangle\right] = \sum_j \frac{|b_j|^2}{M_j}\left[\langle \hat{B}_j^2\rangle - \langle \hat{B}_j\rangle^2\right] =: \sum_j \frac{|b_j|^2}{M_j}V_j.
\end{equation}
Typically the $V_j$ are not known in advance, but if $\hat{B}_j$ are chosen to be unital (by a suitable rescaling of $b_j$) then one can bound $V_j<1$.
One can confirm (e.g. via the use of Lagrange multipliers~\cite{rubin2018application}) that the least number of measurements required to bound $\mathrm{Var}[\langle\mathcal{\hat{H}}\rangle]$ below some error $\epsilon^2$ can be found by choosing $M_j= M|b_j| / \lambda_B$, and that this yields a total number of measurements 
\begin{eqnarray}\label{eq:vqe_measure}
M=\sum_jM_j=\frac{\lambda_B^2}{\epsilon^{2}},
\end{eqnarray}
first mentioned by Ref.~\cite{wecker2015progress}. State tomography is essential in hybrid quantum-classical algorithms such as a VQE. In the VQE, the Hamiltonian is usually expressed as a linear combination of Pauli operators, meaning $\lambda_B=\lambda_Q$.
As the number of measurements is one of the biggest overheads in the practical implementation of VQE, lowering the 1-norm would be highly beneficial.
One way to do so is to add constraints to the Hamiltonian based on the fermionic $n$-representability conditions, which has led to a reduction of up to one order of magnitude for hydrogen chains and the H$_4$ ring molecule~\cite{rubin2018application}.
Another hybrid algorithm that would benefit for a decrease in the 1-norm is the constrained-VQE algorithm, as this would decrease the higher-bound used in the penalty term, thus improving convergence~\cite{kuroiwa2020penalty}.

As the above considerations imply, many of the most competitive algorithms have a strong dependence on 1-norms for various choices of the basis decomposition $\{\hat{B}_j\}$, as shown in Tab.~\ref{tab:scaling_history}.
The 1-norm induced by a Pauli decomposition, $\lambda_Q$ (Eq.~\ref{eq:1norm}), scales with the number of orbitals somewhere in between $\mathcal{O}(N)$ and $\mathcal{O}(N^3)$. This depends on the system, and whether $N$ increases because the number of atoms increases for a fixed number of basis functions per atom, or because the number of basis functions per atom increases while fixing the number of atoms.
As far as we are aware of, this scaling has only been numerically benchmarked for the H$_4$ ring Hamiltonian~\cite{rubin2018application,lee2020even} and one-dimensional hydrogen chains~\cite{lee2020even},
which remain quite far from realistic chemistry problems encountered in enzymatic~\cite{reiher2017elucidating} and catalytic reactions~\cite{von2020quantum}.
In this work, we perform a similar benchmark on a larger set of organic and inorganic molecules,
from hydrogen and carbon chains
to the FeMoco and ruthenium complexes.

The 1-norm used in many of the aforementioned algorithms in Tab.~\ref{tab:scaling_history} takes the following form~\cite{berry2019qubitization,lee2020even}
\begin{eqnarray}\label{eq:lambda}
\lambda = \lambda_T + \lambda_V
\end{eqnarray}
where
\begin{eqnarray}\label{eq:leelambda_T}
\lambda_T = \sum_{pq}^N \left|h_{pq} + \dfrac{1}{2}\sum_{r}^N (2g_{pqrr} - g_{prrq}) \right|
\end{eqnarray}
and
\begin{eqnarray}\label{eq:leelambda_V}
\lambda_V = \dfrac{1}{2}\sum_{pqrs}^N | g_{pqrs} |
\end{eqnarray}
with $\lambda_V \gg \lambda_T$.
Note that the difference between Eqs.~(\ref{eq:leelambda_T}) and (\ref{eq:leelambda_V}) and Eqs.~(A10) of Ref.~\onlinecite{lee2020even} comes from a different convention when writing the electronic Hamiltonian in Eq.~(\ref{eq:el_Ham}), and the fact that $N$ refers to spatial orbitals in this work instead of spin-orbitals.
%\bruno{I think a comparison with Lee et al. should be in the appendix, such that one can really connect our $\lambda_Q$ with the $\lambda_T$ and $\lambda_V$ of there paper (see also my comment at the end of the section):
%To compare with Ref.~\cite{lee2020even} where the one-body and two-body parts are denoted by $T_{pq}$ and $V_{pqrs}$, the following relations can be used:
%\begin{eqnarray}
%h_{pq} &=& T_{pq} - \dfrac{1}{2} \sum_r^N V_{prrq}, \nonumber \\
%g_{pqrs} &=& V_{pqrs}.
%\end{eqnarray}
%}
This norm of the Hamiltonian expressed as a linear combination of unitaries
is used in the database method of Babbush {\it et al.}~\cite{babbush2016exponentially},
the qubitized simulation by the sparse method of Berry {\it et al.}~\cite{berry2019qubitization} (further improved by Lee and coworkers~\cite{lee2020even})
and the qDRIFT protocol of Campbell~\cite{campbell2019random}.
%Note that in general, one would use the standard form of the qubit-Hamiltonian, obtained from the electronic Hamiltonian after a Jordan-Wigner transformation.
%MIn that case, one needs to replace $\lambda_V$ with $\lambda_Q$ in the scalings of Tab.~\ref{tab:scaling_history}.

Further attempts have been made to reduce the number of terms in the Hamiltonian.
For instance, one may perform a low rank decomposition of the Coulomb operator $\hat{V}$
such that the 1-norm of the -- now singly-factorised -- Hamiltonian reads~\cite{berry2019qubitization,lee2020even}
\begin{eqnarray} 
\lambda_{\rm SF}= \dfrac{1}{4}\sum_{\ell=1}^{L} \left(\sum_{p,q}^{N}\left|W_{p q}^{(\ell)}\right|\right)^{2},
\end{eqnarray}
where $W_{pq}^{(\ell)}$ are obtained from a Cholesky decomposition of the $g_{pqrs}$ tensor.
Note that $\lambda_{\rm SF}$ is a higher-bound of $\lambda_V$, and so this decomposition tends to increase the 1-norm.
To improve over this low-rank factorization, one can write the
Hamiltonian in a doubly-factorized form~\cite{von2020quantum} by
rotating the single-particle basis to the eigenbasis of the Cholesky vectors, such that the corresponding 1-norm $\lambda_{\rm DF}$ is much lower than after a single factorization.
One can also use the tensor hypercontraction representation of the Hamiltonian~\cite{lee2020even}.
However, applying the qubitization algorithm on this Hamiltonian directly is not efficient, as its associated 1-norm $\lambda_{\rm THC}$ is even larger than $\lambda_V$.
To bypass this issue, Lee {\it et al.} provided a diagonal form of the Coulomb operator by projection into an expanded non-orthogonal single-particle basis, thus leading to a non-orthogonal THC representation of the Hamiltonian with an associated 1-norm $\lambda_\zeta$ that scales better than any prior algorithm~\cite{lee2020even}.

As readily seen in this section,
the $\lambda$ norm depends on how you represent your Hamiltonian.
However, there is a unique way to write the Hamiltonian as a sum of unique Pauli strings, or equivalently as a sum of unique products of Majorana operators (see Appendix~\ref{appendix:lambda_Q}).
Doing so allows to express the qubit 1-norm $\lambda_Q$ in Eq.~(\ref{eq:1norm}) as a function of the electronic integrals,
\begin{eqnarray}\label{eq:1-norm_Q_pqrs}
    \lambda_Q &=& \lambda_C + \lambda_T + \lambda_V',
\end{eqnarray}
where
\begin{eqnarray}
\lambda_C &=&\left|\sum_p^N h_{pp} + \frac{1}{2}\sum_{pr}^N g_{pprr} - \frac{1}{4}\sum_{pr}^N g_{prrp}\right| \label{eq:lambda_C},\\
\lambda_T &=& \sum_{p q}^N\left|h_{pq} + \sum_r^N g_{pqrr} - \frac{1}{2} \sum_r^N g_{prrq}\right|, \label{eq:lambda_T}\\
\lambda_V' &=& \frac{1}{2}\sum_{p>r, s>q}^N\left|g_{pqrs} - g_{psrq}\right|+ \frac{1}{4} \sum_{pqrs}^N|g_{pqrs}| \label{eq:lambda_V}.
\end{eqnarray}
The first term $\lambda_C$ corresponds to the absolute value of the coefficient of the identity term that emerges when rearranging the Hamiltonian in terms of unique majorana operators, see Eq.~\eqref{eq:Ham_maj1}.
It is invariant under orbital rotations and can be added to the energy as a classical constant together with the nuclear repulsion energy and, if one employs the frozen core approximation (see Appendix \ref{appendix:FrozenCore}), the mean-field energy of the frozen core. 
Thus, this term (apart from being small compared to $\lambda_T$ and $\lambda_V'$) will not be important for quantum algorithms, and we will leave it out of our results, redefining the 1-norm as: $\lambda_Q=\lambda_T + \lambda_V^\prime$.
$\lambda_T$ represents the absolute values of the coefficients of the quadratic term in Majorana operators, and $\lambda_V'$ of the quartic term. 
Notice that $\lambda_V'$ is slightly different from $\lambda_V$ of Ref.~\onlinecite{lee2020even} also given in Eq.~(\ref{eq:leelambda_V}) (which added a slight correction to the one of Ref.~\onlinecite{berry2019qubitization}).
In this work, we add another correction that comes from the swapping of two Majorana operators. 
We confirm that this is the correct form by directly comparing $\lambda_Q$ with the norm of the qubit Hamiltonian after doing a qubit-to-fermion transformation in Eq.~(\ref{eq:1norm}).
Note that $\lambda_V' \leq \lambda_V$ (see appendix \ref{appendix:lambda_Q} for more information).
By expressing the 1-norm of the qubit Hamiltonian in Eq.~(\ref{eq:H_qub}) with respect to the electronic integrals, we can compute its value before doing any fermion-to-qubit transformation, which can be costly for large systems.

% \bruno{I actually think that this equation is an improved version of $\lambda_T + \lambda_V$ ! (If this equation is indeed correct). Indeed, I'm pretty sure we can recover $\lambda_T + \lambda_V$ from the Hamiltonian in Eq.~(\ref{eq:Ham_maj1}), which is certainly equivalent to $H = T' + V'$ in Ref.~\cite{lee2020even}. Then from Eq.~(\ref{eq:Ham_maj1}) to Eq.~(\ref{eq:appendix_lambdaQ}) it seems to me that you improve the representation of the Hamiltonian so that the 1-norm $\lambda_Q$ is reduced compared to $\lambda_T + \lambda_V$.}

\subsection{Localized orbitals}\label{sec:loc}

As discussed previously, different approaches have already been considered to minimize the 1-norm of a given Hamiltonian $\mathcal{\hat{H}}$.
In this work, we choose to tackle the problem from a chemistry point of view, by focusing on the use of orbital transformations (\textit{i.e.} single-particle states rotations). 
More precisely, we investigate the use of orbital localization techniques as a classical pre-optimization method to express the electronic structure Hamiltonian $\mathcal{\hat{H}}$ in a new basis (see Appendix~\ref{appendix:HamTransfo} for details about electronic integrals transformations) that presents natural advantages for quantum computing.
Note that exploiting spatial locality to reduce the 1-norm has already been mentioned in Ref.~\onlinecite{lee2020even}, or to reduce the number of significant integrals in Ref.~\onlinecite{mcclean2014exploiting}.

In computational chemistry, localization schemes represent state-of-the-art orbital-rotation techniques employed in various situations. 
For example, localized orbitals (LOs) are regularly used to alleviate the computational cost of numerical simulations
in
post-HF methods such as second order
M\o{}ller Plesset~\cite{saebo1987fourth,saebo1993local,schutz1999low,weijo2007general},
coupled cluster~\cite{hampel1996local,schutz2001low,tatiana2003local,christiansen2006coupled}, and
multireference methods~\cite{maynau2002direct,angeli2003use,ben2011direct,chang2012multi}.
They can also be used to partition a system in spatially localized subsystems that are treated at different levels of theory~\cite{gomes2012quantum,bennie2015accelerating,hegely2016exact,wouters2016practical,libisch2017embedding,chulhai2017improved,sayfutyarova2017automated,lee2019projection,wen2019absolutely,claudino2019automatic,claudino2019simple,hermes2019multiconfigurational,hermes2020variational}.
In the context of quantum algorithms
for the NISQ era, one may for instance consider performing a calculation with a classical mean-field method 
to produce localized orbitals and using the orbitals localized in the spatial region of interest as a basis
for a calculation with a quantum algorithm.
In the current work, we demonstrate that LOs can also be of a significant help beyond isolating the spatial region that is of chemical interest, by reducing the qubit 1-norm $\lambda_Q$ of the electronic Hamiltonian $\mathcal{\hat{H}}$ after a fermion-to-qubit transformation.

In the following, we introduce the orbital localization schemes considered in this work where the notations
$ \tilde{\psi}_p(\mathbf{r}) $ and $\chi_\mu(\mathbf{r}) $ are used to denote orthogonal LOs and non-orthogonal atomic orbitals (AO), respectively.

\subsubsection{Lowdin orthogonal atomic orbital}

The first approach we investigate to generate LOs is the orthogonalization of atomic orbitals method (OAO). 
In practice, several techniques exist to produce orthogonal AOs~\cite{mayer2002lowdin,szczepanik2013several}.
Here, we focus on L{\"o}wdin's method~\cite{lowdin1950non,carlson1957orthogonalization},
known to generate orthogonal LOs with a shape that is the closest to the original AOs (in the least square sense). 
In practice, orthogonal L{\"o}wdin orbitals $\tilde{\psi}_p(\mathbf{r})$ are built via a linear combination of the $N$ original AOs, 
\begin{equation}
 \tilde{\psi}_p(\mathbf{r})  =   \sum_{\mu=1}^{N}  \chi_\mu(\mathbf{r}) \tilde{{C}}_{\mu p},
\end{equation}
where the orbital coefficient matrix $\tilde{\mathbf{C}}$ takes a very simple form like  
\begin{equation}
    \tilde{\mathbf{C}} = \mathbf{S}^{-1/2},
\end{equation}
and ${S}_{\mu\nu}=\int \ddroit\bfr \chi_\mu^\star(\mathbf{r})\chi_\nu(\mathbf{r}) $ is the overlap matrix encoding the overlap between different non-orthogonal AOs.

From a practical point of view, this method represents one of the simplest and numerically cheapest localization methods, where the computational cost is dominated
by the exponentiation of the overlap matrix and typically scales as $ \mathcal{O}(N^3)$. 
However, this approach is defined with respect to the full AOs' space 
and cannot be straightforwardly applied to the practical case of active space calculations where only a restricted set of molecular orbitals -- formed by linear combination of AOs -- is used. 
%Thus, the OAO scheme can be used in practice if and only if the entire Hilbert space of a given molecular system can be treated in a calculation.
%Which restrains the use of OAO essentially to small-sized molecular systems counting a restricted number of molecular orbitals in total ( \textit{i.e.} $N\sim 20$ at most).

\subsubsection{Molecular orbital localization schemes}

Fortunately, other localization schemes can be used in any circumstances (\textit{i.e.} with or without active space approximation). 
Among the possible approaches, we focus on three of them: the Pipek-Mezey~\cite{pipek1989fast} (PM), Foster-Boys~\cite{foster1960canonical} (FB) and Edmiston-Ruedenberg~\cite{edmiston1963localized} (ER) methods. 
From a practical point of view, these methods fundamentally differ from OAO as they generate LOs out from molecular orbitals and not AOs. 
In practice, all these methods start from a set of orthogonal canonical MOs (CMOs) $\lbrace \phi_s \rbrace_{s=1}^{N}$ obtained with an initial mean-field calculation (\textit{e.g.} Hartree--Fock or density functional theory). 
The CMOs form linear combinations of AOs as
\begin{equation}
    \phi_s(\mathbf{r}) = \sum_\mu^{N} \chi_\mu(\mathbf{r}) {C}_{\mu s},
\end{equation}
with $\mathbf{C}$ the CMO-coefficient matrix. 
A set of LOs $\lbrace \tilde{\psi}_p \rbrace_{p=1}^{N}$ is then generated by applying a unitary transformation matrix $\mathbf{U}$ (with $\mathbf{U^\dag U}=\mathbf{UU^\dag}=\mathbf{1}$) to express each LO as a linear combination of the original CMOs,
\begin{equation}
    \tilde{\psi}_p(\mathbf{r}) = \sum_s^{N} \phi_s(\mathbf{r}) {U}_{sp} ,
    \label{eq:U}
\end{equation}
or, in a more compact matrix form,
\begin{equation}
    \tilde{\mathbf{C}} = \mathbf{C} \mathbf{U},
\end{equation}
where $ \tilde{\mathbf{C}}$ is the LO-coefficient matrix.
In practice, the shape of the LOs is numerically determined by modifying the unitary $\mathbf{U}$ to optimize a cost function $\mathcal{L}$ based on a relevant localization criterion (depending on the scheme considered).
This localization can be prone to many local minima and the orbital-optimization process is usually realized with different numerical techniques (\textit{e.g.} Jacobi rotations \cite{raffenetti1993efficient,barr1975improved}, gradient descent~\cite{lehtola2013unitary}, Newton and trust-region methods \cite{leonard1982quadratically,hoyvik2012trust,sun2016co}, etc.).
In chemistry, the question of how to realize an efficient orbital-optimization process in practice still constitutes an active topic of research which goes way beyond the scope of our manuscript (we refer the interested reader to Ref.~\onlinecite{jensen2017introduction} and references within). 
Let us now focus on the different criteria used in the FB, PM and ER methods together with their respective numerical costs.

In the Foster-Boys scheme~\cite{foster1960canonical}, the localization criterion is the square of the distance separating two electrons $\bfr_{12}^2 = | \bfr_2 - \bfr_1 |^2$,
and the set of LOs is obtained by minimizing the following cost function
\begin{equation}
  \mathcal{L}_\text{FB}   =  \sum_{p}^{N} \iint |\tilde{\psi}_p(\bfr_1)|^2 \bfr_{12}^2 |\tilde{\psi}_p(\bfr_2)|^2    \ddroit \bfr_1 \ddroit \bfr_2,
  \label{eq:FB}
\end{equation}
which represents the average value of the distance criterion over the set of orbitals to be optimized. 
In practice, a single estimation of the cost function $\mathcal{L}_\text{FB}$ requires
a decomposition in the AO basis that generates five nested sums, thus leading to a  scaling of $\mathcal{O}(N^5)$. 
However, specific manipulations (see Refs.~\onlinecite{foster1960canonical} and \onlinecite{jensen2017introduction}) can reduce the total computational cost of the Foster-Boys method to $\mathcal{O}(N^3)$ multiplied by the number of times $\mathcal{L}_\text{FB}$ is called (which intrinsically depends on the optimization algorithm considered in practice).
Note that extensions of the Foster-Boys scheme exist and are based on the orbital variance
~\cite{jansik2011local} or the fourth moment method~\cite{hoyvik2012orbital}, resulting in more localized orbitals especially for basis sets augmented by diffuse functions.

In the case of the Edmiston-Ruedenberg method, the inverse distance $1/\bfr_{12}$ (proportional to the two-body electronic repulsion operator, see Eq.~(\ref{eq:eri}))
is used as a criterion, and the LOs are obtained by maximizing
\begin{equation}
  \mathcal{L}_\text{ER}  =  \sum_{p}^{N} \iint |\tilde{\psi}_p(\bfr_1)|^2 \frac{1}{\bfr_{12} }  |\tilde{\psi}_p(\bfr_2)|^2   \ddroit \bfr_1 \ddroit \bfr_2.
  \label{eq:ER}
\end{equation}
The computational cost of a single estimation of $\mathcal{L}_\text{ER}$ scales as $\mathcal{O}(N^5)$ (multiplied by the number of times $\mathcal{L}_\text{ER}$ is called),
though it can be reduced up to $\mathcal{O}(N^3)$ by the use of density fitting
or the Cholesky decomposition.

Finally, the Pipek-Mezey scheme adopts a very different point of view. PM orbitals are generated by maximising the Mulliken charge of each orbital~\cite{pipek1989fast}
\begin{equation}\label{eq:L_PM}
 \mathcal{L}_\text{PM}   =  \sum_{A=1}^{N_A} \sum_p^N (Q_A^p)^2,
\end{equation}
where
\begin{eqnarray}
Q_A^p = \sum_{\mu \in A} \sum_{\nu}^{N} \tilde{{C}}_{\mu p} {S}_{\mu\nu} \tilde{{C}}_{\nu p}
\label{eq:PM}
\end{eqnarray}
is the contribution of
orbital $p$ to the Mulliken charge of atom $A$.
In this case, the numerical cost of one call of $ \mathcal{L}_\text{PM} $ essentially depends on the triple sums over the molecular and atomic orbitals in Eqs.~(\ref{eq:L_PM}) and (\ref{eq:PM}).
As a results, the computational cost of the global PM method scales as $\mathcal{O}(N^3)$ (multiplied by the number of times $\mathcal{L}_\text{PM}$ is called).
Note that the traditional PM method is ill-defined since Mulliken charges are basis set sensitive and do not have a basis set limit.
Various partial charges estimates can be used instead of Mulliken charges~\cite{lehtola2014pipek}, such as intrinsic orbitals that are basis set insensitive and lead to a
cheaper and better behaved localization procedure than PM 
localization~\cite{knizia2013intrinsic,senjean2021generalization}.

In summary, the construction of OAOs is the cheapest approach but can't be used when considering active space. Somewhat more expensive are PM and FB which share an equivalent scaling, and finally ER which is the most computationally demanding method.

\subsection{1-Norm orbital-optimization}\label{sec:1-norm-opt}

One of the main contribution of this paper is the use of an orbital-optimization (OO) process specifically dedicated to the minimization of the qubit 1-norm $\lambda_Q$. 
To proceed, we introduce a unitary operator
\begin{equation}
    \mathbf{U}^\text{OO} = e^{-\mathbf{K}} 
\end{equation}
with an anti-hermitian generator
\begin{equation}
    \mathbf{K}^T = -\mathbf{K} =
    \begin{pmatrix}
    0 & K_{12} & K_{13} & \ldots & K_{1N} \\
    -K_{12} & 0 & K_{13} & \ldots & K_{2N} \\
    -K_{13} & -K_{13} & 0 & \ldots & K_{3N} \\
    \vdots & \vdots & \vdots & \ddots & \vdots \\
    -K_{1N} & -K_{2N} & -K_{3N}  & \ldots & 0
    \end{pmatrix}.
\end{equation}
This operator is then used to transform a reference MO basis into a new basis denoted by $\lbrace \phi'_q \rbrace$ with $\mathbf{C}'= \mathbf{C} \mathbf{U}^\text{OO}$,
for which the optimal orbitals are obtained by minimizing the following cost function,
\begin{equation}\label{eq:CF_1norm}
    \mathcal{L}_{OO} = \lambda_Q(\lbrace \phi'_q \rbrace),
\end{equation}
by varying the $N(N-1)/2$ off-diagonal independent elements of the matrix $\mathbf{K}$. This optimization can
be carried out for a reference MO-basis that consists of canonical Hartree-Fock orbitals, but could equally
well be carried out on a subspace of the full MO-basis, e.g. only the set of fractionally occupied orbitals in an active
space type calculation, or only a subset of localized orbitals resulting from a preliminary localization using
one of the methods discussed above.

In practice, to realize the OO of the 1-norm, we choose to use `brute-force' optimization algorithms which autonomously estimate the local derivatives of the cost function (\textit{i.e.} gradients and hessians). This choice is motivated by the expression of $\lambda_Q$ in Eq.~(\ref{eq:1-norm_Q_pqrs}) that contains many absolute values and for which analytical derivatives are clearly non-trivial to estimate. Numerically, the main bottleneck of the OO method is linked to the repetitive transformation of the two-electron integrals realized at each step of the process. As a result, the core algorithm scales as $\mathcal{O}(N^5)$ (similarly to the ER localization scheme). However, with the computational effort deployed to estimate the numerical gradient and hessian of the complicated cost function (given in Eq.~(\ref{eq:CF_1norm})), this scaling should increase more especially when treating large systems.

\section{Computational details}\label{sec:comp}

%The goal is to calculate $\lambda_Q$ for various molecules, either in the full or in an active space. 
The geometries of the small systems considered in this work were optimized using the ADF program of the Amsterdam Modeling Suite (AMS)~\cite{tevelde2001chemistry} with a quick universal force-field (UFF) optimization, sufficient for our purposes. These geometries are provided in the Supplementary Material~\cite{supp_mat}.
The first geometry in Ref.~\onlinecite{reiher2017elucidating} with a charge of +3 is used for FeMoco and the geometries in Ref.~\onlinecite{von2020quantum} (all with a charge of +1) were used for the Ruthenium metal complexes.
The electronic integrals of the Hamiltonian were computed using the restricted Hartree--Fock method from the PySCF package~\cite{sun2017pyscf}.
For large molecules, the frozen-core approximation was invoked according to Appendix~\ref{appendix:FrozenCore}. 
All the localization schemes used to transform the Hamiltonian were already implemented in the PySCF package.
For our 1-norm orbital-optimization scheme, the SLSQP optimizer was used from the SciPy package~\cite{virtanen2020scipy} and the python version of the L-BFGS-B optimizer of the optimParallel package~\cite{gerber2019optimparallel}.
The choice of these optimizers has been motivated by their capacity to automatically approximate gradients (and Hessians) for minimization. This represents an interesting tool when no evident analytical gradient/Hessian can be determined as for the 1-norm.
A fast algorithm to estimate this $\lambda_Q$ [Eq.~(\ref{eq:1-norm_Q_pqrs})] was implemented in the OpenFermion package \cite{mcclean2020openfermion}, allowing users to calculate 1-norms of large molecular Hamiltonians without employing expensive fermion-to-qubit mappings.

\section{Results}\label{sec:results}

In this section, we study the scaling of $\lambda_Q$ with respect to the number of orbitals. First, we fix the basis set and increase the number of atoms, in the spirit of Refs.~\onlinecite{rubin2018application} and \onlinecite{lee2020even}. Concerning the orbital localization schemes, we allow rotations between the active occupied and virtual spaces as this can lead to a better localized orbital basis.

% As an illustration, see the example in Fig.~\ref{fig:jacobi_example} where the occupied $\pi$-bonding orbital mixes with the virtual $\pi^*$-antibonding orbital for the formaldimine molecule.
% By rotating those two orbitals by a Jacobi rotation, we move from a maximal value of $\lambda_Q$ for $\theta = 0$ (corresponding to the original CMOs)
% to a minimal value for $\theta = \pi/4$ (corresponding to  the new LOs).

% \begin{figure}
%     \centering  
%     \includegraphics[width=8cm]{MO_LOC_1norm.pdf}
%     \caption{Moving between the CMOs, corresponding to the occupied $\pi$-bonding and virtual $\pi^*$-antibonding orbitals of \ch{HNCH2}, to the LOs by performing a Jacobi rotation $\mathbf{U}(\theta) = \begin{pmatrix}
%     \cos{\theta} & \sin{\theta}\\
%     -\sin{\theta} & \cos{\theta}
%     \end{pmatrix}$.}
%     \label{fig:jacobi_example}
% \end{figure}

Then, we benchmark the value of $\lambda_Q$ for several different chemical systems and active space sizes, using the different localization schemes and our 1-norm orbital-optimized scheme.
When considering an active space, we always choose our active space based on the CMOs by considering the $N_{\rm act}$ orbitals around the Fermi level (\textit{i.e.} around the highest occupied and lowest unoccupied molecular orbitals).
We localize only \textit{inside} the active space (i.e. the localizing unitary in Eq.~\eqref{eq:U} only has indices corresponding to active orbitals)
such that the subspace spanned by the active space remains invariant under the unitary rotations.
Although this means that the expectation values of observables remain the same inside the active space when an exact solver is used,
one can converge to different expectation values when approximate solvers are considered such as the truncated unitary coupled cluster ansatz (that becomes exact if not truncated~\cite{evangelista2019exact}).
However, rotating from the CMO to the LMO basis will not necessarily deteriorate the results of the (approximate) simulation,
as LMOs have significant importance in
local correlation treatments in
post-HF methods like second order
M{\o}ller Plesset~\cite{saebo1987fourth,saebo1993local,schutz1999low,weijo2007general},
coupled cluster~\cite{hampel1996local,schutz2001low,tatiana2003local,christiansen2006coupled},
embedding approaches~\cite{gomes2012quantum,bennie2015accelerating,hegely2016exact,wouters2016practical,libisch2017embedding,chulhai2017improved,sayfutyarova2017automated,lee2019projection,wen2019absolutely,claudino2019automatic,claudino2019simple,hermes2019multiconfigurational,hermes2020variational}, and
multireference methods~\cite{maynau2002direct,angeli2003use,ben2011direct,chang2012multi} on classical computers.
Hence, using localized orbitals could lead to similar advantages in quantum computing simulations, and could also inspire new ansatz proposals based on embedding strategies. This is left for future work.

Finally, to study the scaling of $\lambda_Q$ for large molecules, we increase the size of the active space $N_{\rm act}$ while fixing the basis set and the number of atoms.

\subsection{Hydrogen and alkane chains: scaling of the 1-norm by increasing the number of atoms}\label{sec:HandC}

\begin{figure}
\centering
    \topinset{\scalebox{1.5}{\bfseries{(a)}}}{\includegraphics[width=.9\linewidth]{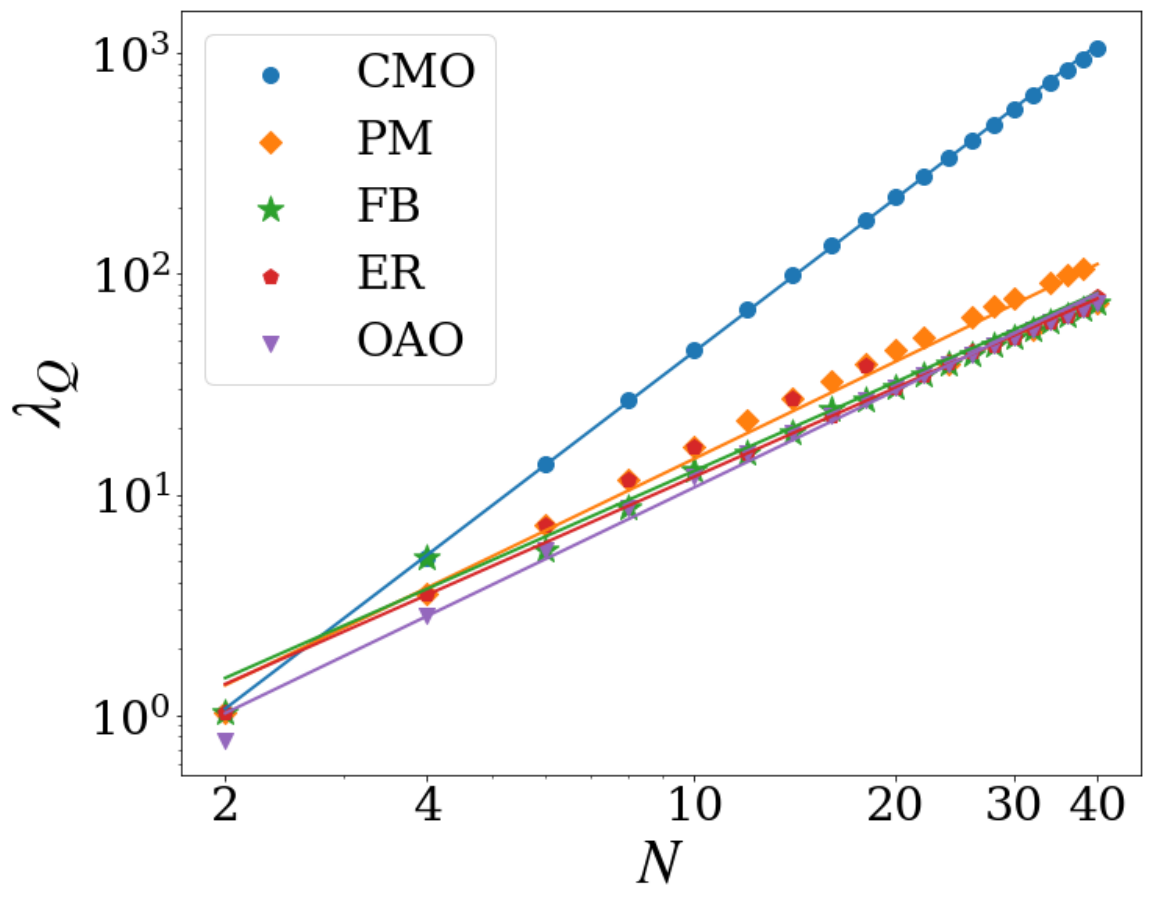}}{-.01in}{-1.45in}
    \topinset{\scalebox{1.5}{\bfseries{(b)}}}{\includegraphics[width=.9\linewidth]{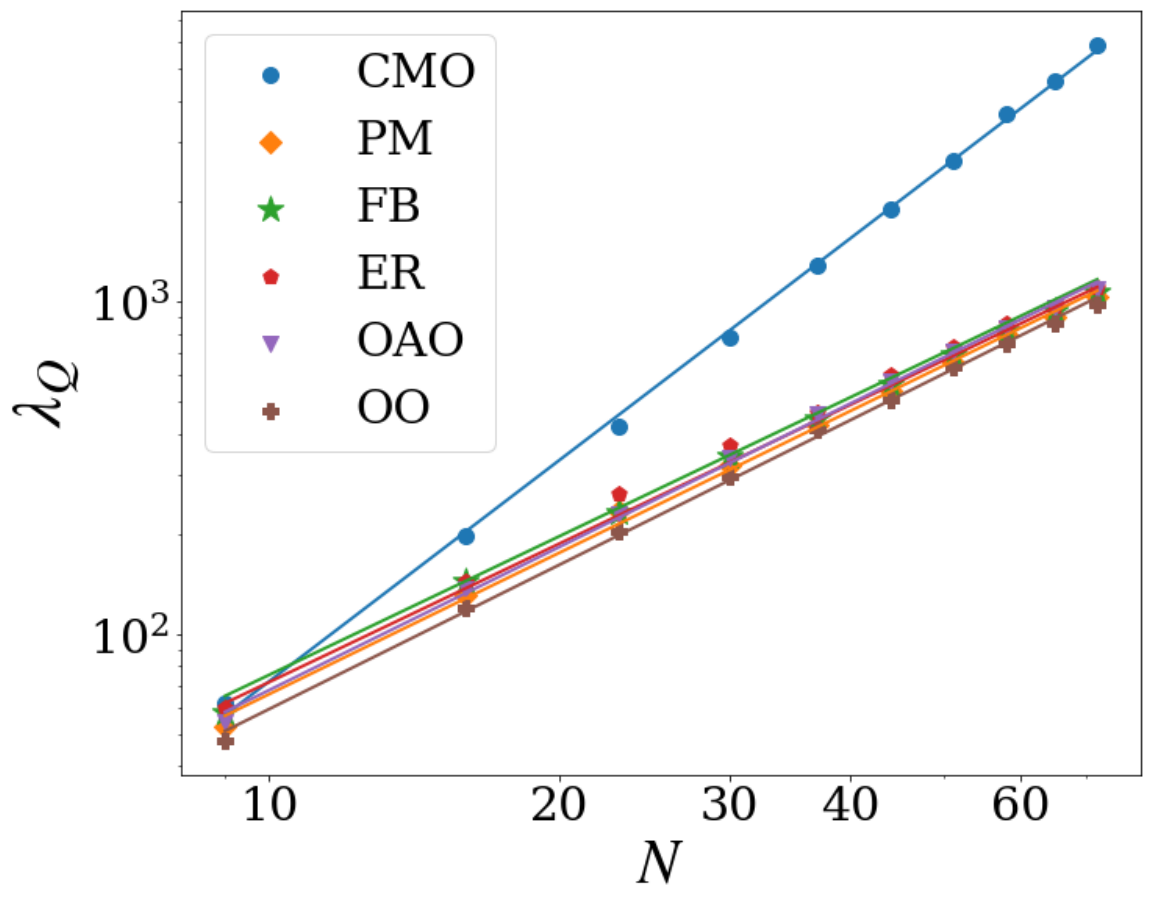}}{0.in}{-1.45in}
    \caption{ 
    The 1-norm $\lambda_Q$ (in Hartree) with respect to {\bfseries{(a)}} the number of atoms in a linear hydrogen chain (same as the number of orbitals) and {\bfseries{(b)}} the number of orbitals in a linear alkane chains (ethane, propane, butane, etc. up to decane), for different orbitals in the STO-3G basis. The constant term in the Hamiltonian is ignored. Assuming a polynomial scaling of $\lambda_Q = \mathcal{O}(N^\alpha)$, we fit $\log{\lambda_Q}=\alpha \log{N} +\beta$ and show the plot with log-log axes. The fitting and regression coefficients are given in Tab.~\ref{tab:R_2values}.}
    \label{fig:linear_chains}
\end{figure}

In this section, we investigate the performance of the localization schemes together with our brute-force OO method, studying the scaling of $\lambda_Q$ with respect to the number of orbitals by increasing the number of atoms in small systems.
Inspired by Refs.~\onlinecite{mcclean2014exploiting} and \onlinecite{rubin2018application}, 
we consider linear chains of hydrogens with a spacing of $r = \SI{1.4}{\angstrom}$ and linear alkane chains (for which the geometry has been optimized).
Results are shown in Fig.~\ref{fig:linear_chains} and Tab.~\ref{tab:R_2values}.

The orbital-optimizer method is not included for the linear hydrogen chains, because the gradient-estimating optimizer has trouble finding the approximate gradient in the first step of the optimization. The reason for this is that the localization schemes already come so close to the optimal solution that the optimizer estimates the gradient to be zero or positive in each direction.
% Hydrogen chains have a high degree of symmetry resulting in too much freedom of the unitary basis transformation to have any effect on $\lambda_Q$.
The STO-3G basis results in the localized orbitals resembling just 1s atomic orbitals on every nucleus which are very localized and therefore have a minimal 1-norm. As linear hydrogen chains are very artificial systems and give limited information on the scaling of $\lambda_Q$ for actual interesting chemical systems, we thought the results of the conventional localization schemes to be sufficient here.

% \begin{figure}
%      \centering
%      \begin{subfigure}[b]{.9\linewidth}
%          \centering
%          \includegraphics[width=.9\linewidth]{h_chain_loglog.pdf}
%         %  \caption{$y=x$}
%          \label{fig:h_chain}
%      \end{subfigure}
%      \\
%      \begin{subfigure}[b]{.9\linewidth}
%          \centering
%          \includegraphics[width=.9\linewidth]{c_chain_OO.pdf}
%         %  \caption{$y=3sinx$}
%          \label{fig:c_chain}
%      \end{subfigure}
%         \caption{The 1-norm $\lambda_Q$ (in Hartree) with respect to the number of atoms in a linear hydrogen chain (same as the number of orbitals) seen in the upper panel and the number of orbitals in a linear alkane chains (ethane, propane, butane, etc. up to decane) in the lower panel, for different orbitals in the STO-3G basis. The constant term in the Hamiltonian is ignored. Assuming a polynomial scaling of $\lambda_Q = \mathcal{O}(N^\alpha)$, we fit $\log{\lambda_Q}=\alpha \log{N} +\beta$ and show the plot with log-log axes. The fitting and regression coefficients are given in Tab.~\ref{tab:R_2values}.}
%         \label{fig:linear_chains}
% \end{figure}

\begin{table}
    \centering
    \begin{tabular}{|c||c|c|c|}
    \hline
    \thead{Hydrogen chain}     & \thead{$\alpha$} &  \thead{$\beta$} & \thead{$R^2$} \\
    \hline
    CMO    &  2.31  &  -1.54  & 0.9990 \\
    \hline
    PM     &  1.47  &  -0.70  & 0.8832 \\
    \hline
    FB     &  1.34  &  -0.61  & 0.9967 \\
    \hline
    ER     &  1.34  &  -0.54  & 0.9643 \\
    \hline
    OAO    &  1.46  &  -0.99  & 0.9846 \\
    \hline
    \end{tabular}
    \vadjust{\vskip 3mm \vskip 0pt}
    \begin{tabular}{|c||c|c|c|}
    \hline
    \thead{Alkane chain}     & \thead{$\alpha$} &  \thead{$\beta$} & \thead{$R^2$} \\
    \hline
    CMO    &  2.21  &  -0.82 & 0.9978 \\
    \hline
    PM     &  1.41  &  0.93  & 0.9965 \\
    \hline
    FB     &  1.38  &  1.08  & 0.9976 \\
    \hline
    ER     &  1.39  &  1.13  & 0.9926 \\
    \hline
    OAO    &  1.43  &  0.92  & 0.9954 \\
    \hline
    OO     &  1.44  &  0.76  & 0.9964 \\
    \hline
    \end{tabular}
    \caption{Fitting and regression coefficients from Fig.~\ref{fig:linear_chains}.}
    \label{tab:R_2values}
\end{table}

As readily seen,
% in Fig.~\ref{fig:linear_chains} and Table~\ref{tab:R_2values},
all localization schemes perform comparably well in reducing the 1-norm
compared to canonical molecular orbitals.
Indeed, the scaling of the 1-norm decreases significantly from $\mathcal{O}(N^{2.31})$
to $\mathcal{O}(N^{1.34})$ for the hydrogen chains, and from $\mathcal{O}(N^{2.21})$
to $\mathcal{O}(N^{1.38})$ for the alkane chains.
In terms of concrete values, for the largest $N$ studied in Fig.~\ref{fig:linear_chains}, we can see a reduction of $\lambda_Q$ by more than a factor 13 for the hydrogen and 5 for the alkane chains, when passing from the CMO basis to the LO ones. 
In the alkane chains, we do not see any significant distinction between the different schemes. While $\alpha$ for the orbital-optimization scheme is slightly higher than the best localization scheme, it is not so informative here as the $\lambda_Q$ associated to the OO has the lowest absolute value for any orbital space.
% This could be because the orbital-optimizer is harder to converge and is likelier to converge to a local minimum for larger systems -- even more so than conventional localization schemes.
In the hydrogen chains, it seems that the Pipek-Mezey scheme is slightly less efficient. 
Surprisingly, the atomic orbital based Löwdin orthogonalization method do not provide any further improvement over other localized orbitals, which might be due to the simplicity of the systems for which localization schemes can generate extremely localized orbitals.

\subsection{Benchmarking $\lambda_Q$ for a variety of molecules and active spaces}\label{sec:benchmark}

In order to benchmark the impact of the localization schemes on the 1-norm,
we calculated $\lambda_Q$ for a variety of molecules and active spaces.
Whenever we consider the full space of orbitals, we include the value of $\lambda_Q$ obtained from the Hamiltonian expressed in the OAO basis.
When an active space is considered,
we use the frozen-core approximation and we localize the orbitals inside the active space.
The resulting $\lambda_Q$ values are tabulated in Tab~\ref{CAS_table}, where the lowest ones obtained from standard localization schemes for each system are in bold.
Apart from the different localization schemes, results of our brute force optimizer are also shown. 
Note that we start the optimization  with already localized orbitals, such that it always gives a lower 1-norm than the best localization scheme.
On most molecules, we had the best experience with the L-BFGS-B optimizer. Unfortunately, this optimizer was not able to estimate the gradient of $\lambda_Q$ w.r.t. $\mathbf{U}^\text{OO}$ on the smallest molecules (\ch{H2} and \ch{LiH}), which may be due to infinitesimal changes in $\lambda_Q$ when changing $\mathbf{U}^\text{OO}$. This why we employed the SLSQP optimizer here.

%\makeatletter\onecolumngrid@push\makeatother
\begin{table*}
\centering
\begin{tabular}{|l|l||l|l|l|l|l|l|l|}
\hline
\thead{ Molecule }&\thead{Active space}&\thead{CMO}&  \thead{PM}    & \thead{FB}     & \thead{ER}      & \thead{OAO}     &\thead{Optimizer}& \thead{\% Reduction}\\ 
\hline
\ch{H2}           &  Full space (2,10) &  101      &   135          &     116        &   \textbf{93}  &      103        &  90$^{(a)}$    & 10.9\% \\ 
\hline
\ch{LiH}          &  Full space (4,19) &   185     &   177          &   190          &   178           &   \textbf{153}  &   $134^{(a)}$   &  27.6\%\\ 
\hline
\ch{H2O}          &  Full space (10,24)&   717     &   678          &    710         &   662           &   \textbf{616}  &  $576^{(b)}$    & 19.7\% \\ 
\hline
\ch{HLiO}         &  Full space (12,33)&   993     &   787          &    831         &   \textbf{768}  &   788           &  $668^{(b)}$    & 32.7\%\\ 
\hline
\ch{H2CO}         &  Full space (16,38)&   1792    &   1419         &     1417       &    1343         &    \textbf{1327}&  $1101^{(b)}$   &  38.6\%\\ 
\hline
\ch{HNCH2}        &  Full space (16,43)&   3096    &   1813         &    1670        &   \textbf{1588} &      1645       &  $1240^{(b)}$   &  60.0\%\\ 
\hline
\ch{C3H6}         &  (24,45)           &   2316    &   1223         &    1145        &   \textbf{1137} &      N/A        &  $995^{(b)}$    &  57.0\%\\
\hline
\ch{C4H6}         &  (30,45)           &   1760    &   \textbf{974} &    1054        &   1048          &      N/A        &  $812^{(b)}$    &  53.9\%\\
\hline
\ch{C5H8}         &  (38,45)           &   1821    &   818          &    801         &   \textbf{799}  &      N/A        &  $698^{(b)}$    &  61.7\%\\
\hline
\ch{HNC3H6}       &  (32,50)           &   3796    &   1493         &    1244        &    \textbf{1232}&      N/A        & $1085^{(b)}$    &  71.4\%\\ 
\hline
\ch{HNC5H10}      &  (48,50)           &   2632    &   1098         &   \textbf{989} &   1002          &      N/A        &   $842^{(b)}$   &  68.0\%\\ 
\hline
\ch{HNC7H14}      &  (50,50)           &   2610    &   751          &   \textbf{705} &   707           &      N/A        & $616^{(b)}$     &  76.4\%\\ 
\hline
\end{tabular}
\caption{Values of $\lambda_Q$ for relatively small test molecules in the cc-pVDZ basis-set.
The active space size is indicated as $(n_e,N)$, where $n_e$ is the number of electrons in $N$ the number of spatial orbitals. 
The lowest 1-norm obtained from standard localization schemes for each system are in bold.
Superscripts ($a$) and ($b$) refer to the use of the SLSQP and the L-BFGS-B optimizer used for in our 1-norm orbital-optimization scheme (denoted as ``Optimizer'' in the table), respectively. 
The rightmost column shows the percentage of reduction obtained from the 1-norm-optimized orbitals compared to the CMOs.}
\label{CAS_table}
\end{table*}

\begin{figure}
\centering
    \includegraphics[width=\columnwidth]{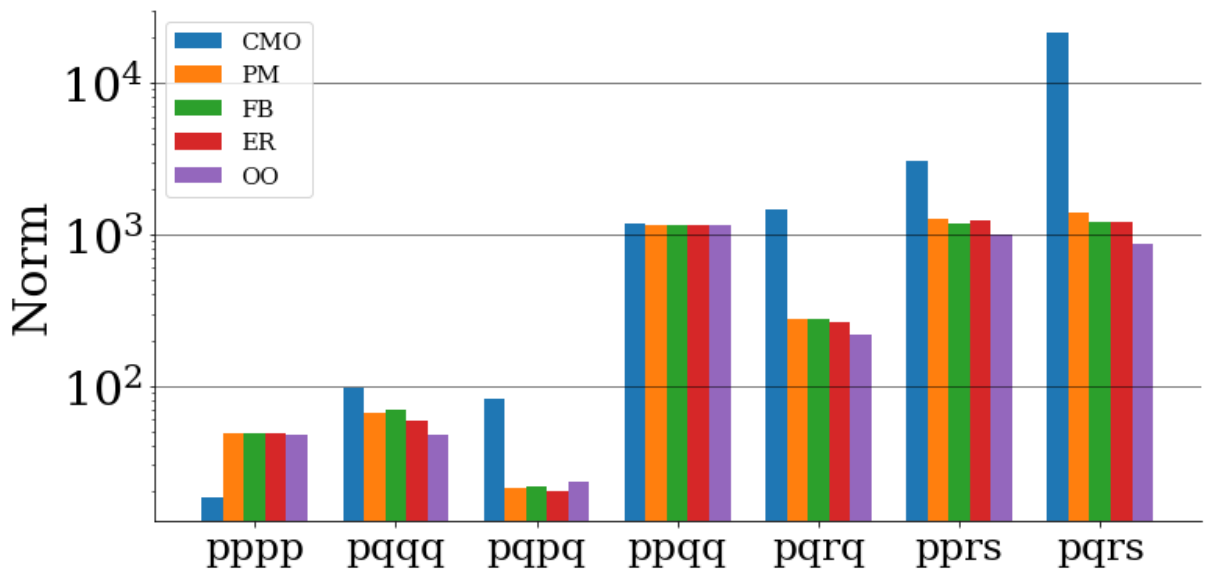}
\caption{
    % \bruno{Suggestion: remove the numbers and set this figure in one-column format. Add grids for a guide to the eye. The numbers can be discussed in the main text (so keep this figure somewhere for us). To me, the precise numbers are not needed. The important part is to see that it decreases the 1-norm by more than an order of magnitude. Then, change size of legend and just use the same as in previous figures: CMO, PM, FB, ...}
    % \saad{Use the same style of font as in the previous figures to be coherent. Furthermore, as most of the time the columns are bigger on left than on the right change the rotation to avoid overlap. Use a grid also on the Y-axis and maybe try to see if you can write bigger numbers at the top of the columns (maybe we could actually get rid of these numbers)} 
    (Color) Sum of absolute values of different parts of the two-electron integral tensor on a logarithmic scale (see text for further details on the notations). It shows the second stable Ruthenium intermediate in a minimal ANO-RCC basis for an active space of 100 electrons in 100 spatial orbitals. \label{fig:ferm_ints}}
\end{figure}

In order to identify the origin of the significant reduction of the 1-norm and its scaling, we focus on the transformation of the two-electron integrals defined in Eqs.~(\ref{eq:eri})
with the following permutational symmetries (for real-valued integrals)
\begin{eqnarray}\label{eq:intsym}
&&(pq|rs) = (pq|sr) = (qp|rs) = (qp|sr) \nonumber \\
=&&
(rs|pq) = (rs|qp) = (sr|pq) = (sr|qp) ,
\end{eqnarray}
and we divide the indices of this tensor in seven separate classes:
(1) pppp,
(2) pqqq,
(3) pqpq,
(4) ppqq,
(5) pqrq,
(6) pprs,
(7) pqrs,
where the indices p,q,r and s are all different.
The above notations are used in Fig.~\ref{fig:ferm_ints} and correspond to 
the sum of the absolute values of the electronic integrals associated to these indices [and the ones related by symmetries in Eq.~(\ref{eq:intsym})], e.g. 
\begin{eqnarray}
\text{pqrq} &\equiv& \sum_{p \neq q \neq r } |g_{pqrq}| + |g_{pqqr}| + |g_{qprq}| + |g_{qpqr}|\nonumber\\
&&= \sum_{p\neq q \neq r} 4 |g_{pqrq}|\nonumber \\
\text{pqrs}& \equiv& \sum_{p\neq q \neq r \neq s} |g_{pqrs}|
\end{eqnarray}
for the cases (5) and (7), respectively.
These contributions can give us some insight as to how the localized orbitals affect the norm of the diagonal and off-diagonal parts of the tensor,
and are represented in Fig.~\ref{fig:ferm_ints} for
% the FeMoco and 
the second stable intermediate Ruthenium complex in different orbital bases.
As readily seen in this diagram,
localization schemes do not lead to a uniform reduction of the norm of every term, but tends to maximize some of them (the pppp term), minimize others (the pqrs term) and leaves others relatively unchanged (the ppqq term).
However, the pqrs contribution clearly dominates in the CMO basis, being more than one-order of magnitude larger than any of the other terms. Especially the orbital-optimizer can give an even better reduction of this term.
This explains why the localization schemes can be used to reduce the 1-norm, as
reducing the pqrs norm by one order of magnitude (as seen in Fig.~\ref{fig:ferm_ints}) will play a much more important role than increasing the pppp one by the same order of magnitude. As the magnitude of these integrals depends on the overlap densities $\psi_p(\mathbf{r})\psi_q(\mathbf{r})$ and $\psi_r(\mathbf{r})\psi_s(\mathbf{r})$, it is clear that localization will
reduce the number of numerically significant contributors. For an extended
system, integrals in which p and q as well as r and s are localized on
the same atoms are expected to dominate. This explains the observation that the 
pqrs norm becomes comparable to the ppqq and pprs terms when employing
localized orbitals.

\subsection{Effect of increasing the size of the active space on $\lambda_Q$}\label{sec:lambda_scaling}

Let us now investigate how $\lambda_Q$ scales with respect to the number of active orbitals for large and complex systems.
We selected test-molecules based on two important recent papers in the field of quantum computing for quantum chemistry: FeMoco -- an iron-molybdenum complex that constitutes the active site of a MoFe protein which plays an important role in nitrogen fixation~\cite{reiher2017elucidating} -- and all the stable intermediates and transition states of the catalytic cycle presented in Ref.~\onlinecite{von2020quantum}.
This catalytic cycle describes the binding and transformations of carbon dioxide \ch{CO2}, a molecule with infrared absorption properties that makes it a potent and the most important greenhouse gas.

\begin{figure}
    \centering
    \topinset{\scalebox{1.5}{\bfseries{(a)}}}{\includegraphics[width=\linewidth]{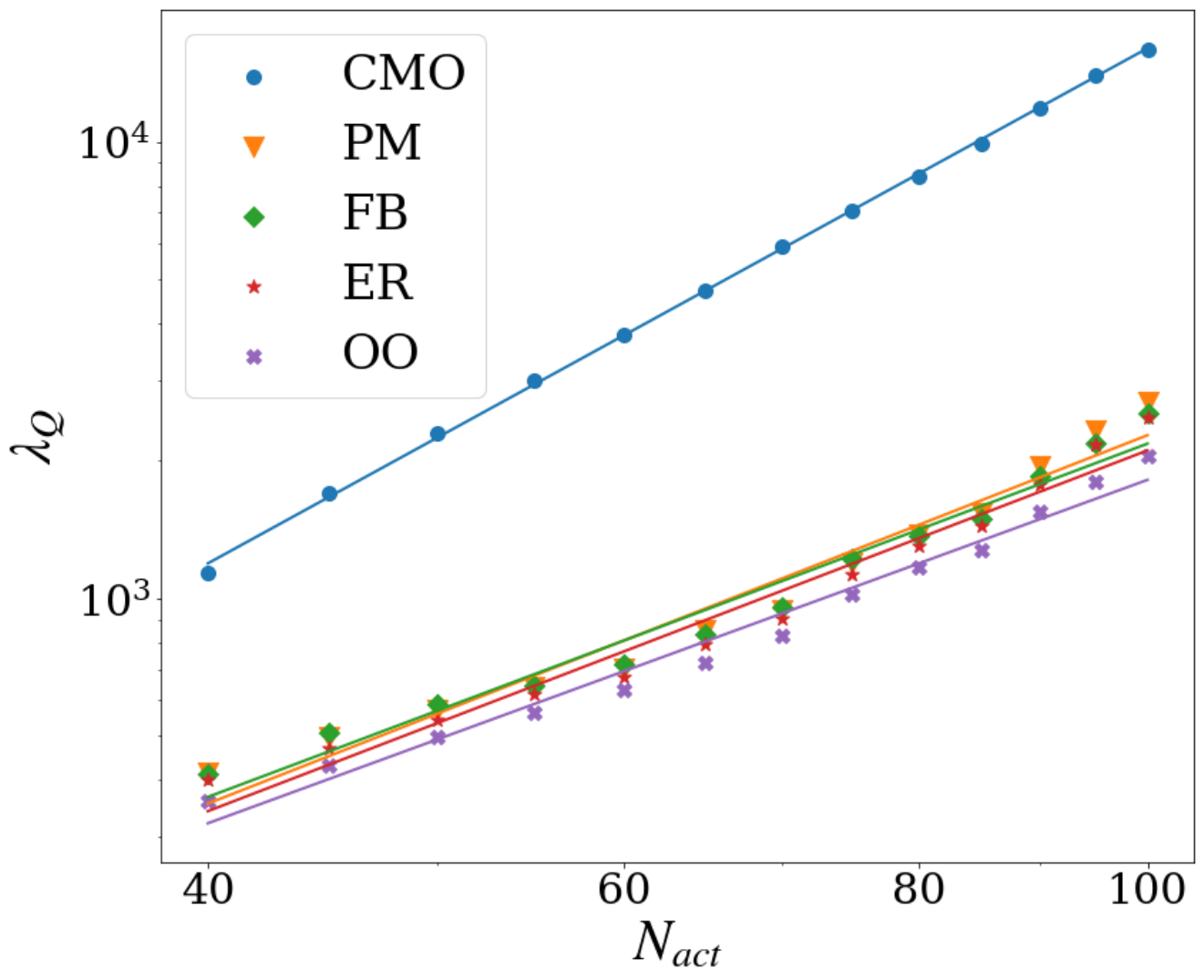}}{-.01in}{-1.45in}
    \topinset{\scalebox{1.5}{\bfseries{(b)}}}{\includegraphics[width=\linewidth]{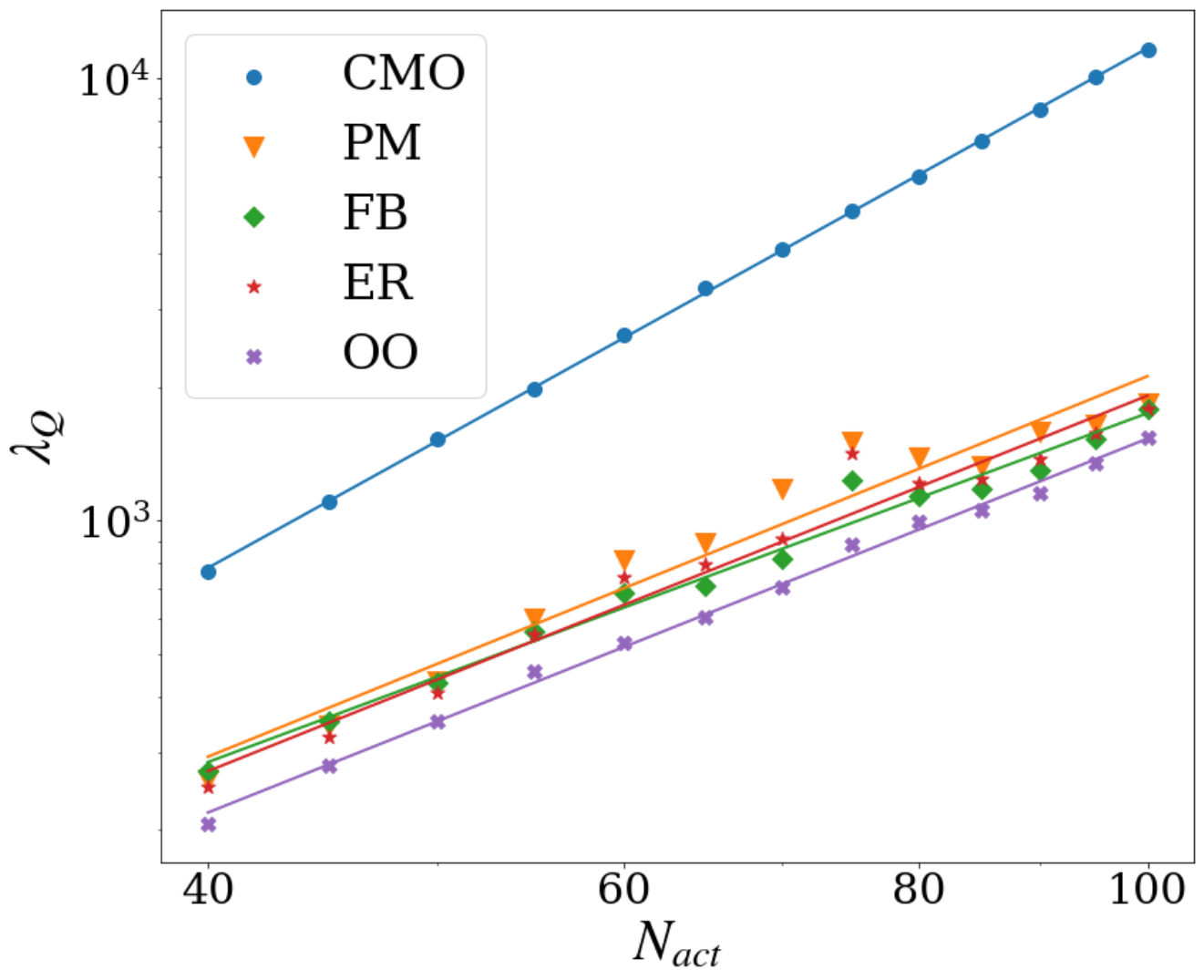}}{-0.12in}{-1.45in}
    \caption{Scaling of $\lambda_Q$ (in Hartree) with respect to the size of the active space for different orbitals for {\bfseries{(a)}} FeMoco and {\bfseries{(b)}} the Ruthenium transition state II-III in a minimal ANO-RCC basis. The constant term in the Hamiltonian is ignored. Assuming polynomial scaling of $\lambda_Q = \mathcal{O}(N_{\rm act}^\alpha)$ we fit $\log{\lambda_Q}=\alpha \log{N_{\rm act}} +\beta$ and show the plot with log-log axes. The fitting parameters and regression coefficients are given in Table~\ref{tab:scaling}. \label{fig:Ru_detail}}
\end{figure}

For the calculations to remain doable in a reasonable amount of time, the ANO-RCC minimal basis (with scalar relativistic corrections) is used. This should be sufficient for the current purpose of studying 1-norm reductions within an active space, but we note that larger basis sets will be required
to reach chemical accuracy~\cite{Halkier1998}.
The scaling of $\lambda_Q$ with respect to the number of active orbitals $N_{\rm act}$ for different orbital bases are reported in details in Fig.~\ref{fig:Ru_detail}.
Comparing the scaling reported in this section and the ones obtained in Sec.~\ref{sec:HandC}, we see that the scaling is larger when increasing the number of active orbitals than when increasing the number of atoms, as expected.
However, the impact of localized orbitals on the 1-norm is similar in both cases, as it reduces the scaling by almost one order of magnitude.
We also do not see any significant distinction between the 1-norm associated to the different conventional localization schemes. The brute-force orbital-optimizer however, gives a consistent 10-25\% improvement over the conventional localization schemes. In the largest active spaces considered ($N_{\rm act} \geq 85$) the orbital-opimization scheme can converge very slowly due to the need of 4-index transformations at every step, combined with the non-continuous and complicated landscape of $\lambda_Q$. Because of this, the user is usually forced to cut off the optimizer at some point, or choose a high threshold of convergence. In spite of this, we got similar good results in the largest active spaces we considered, indicating that the optimizer is not limited to a specific active space size.

The scaling results for all large molecules are summarized in Table~\ref{tab:scaling}. 
The scaling lies in the order of $\mathcal{O}(N^{2.6})$ to $\mathcal{O}(N^{2.9})$ when using CMOs, while it lies around $\mathcal{O}(N^{2})$ or slightly higher when using LOs. One gets a bit of a misleading picture looking at just the scaling of the orbital-optimizer, as it has the lowest absolute value of $\lambda_Q$ at every point. The cause is that the cost function of the orbital-optimizer has more local minima and is harder to converge compared to conventional localization schemes.

% For the localization schemes it is usual that the bigger the active space is, the more $\lambda_Q$ can be reduced compared to CMOs.
% This can be rationalized as follows.
% When the active space increases, it may involve orbitals localized on atoms which have negligible overlap with the other orbitals.
% Another possibility is that the additional orbitals may have a positive effect on the localization procedure (like the addition of a $\pi^*$-antibonding orbital in the example illustrated in Fig.~\ref{fig:jacobi_example}).
% In other words, there are more degrees of freedom in a larger active space to perform a better orbital localization. For the biggest active space chosen, an active space of 100 electrons in 100 orbitals, $\lambda_Q$ is reduced by an order of magnitude. This illustrates the potential of using widely available and computationally cheap classical pre-optimizations, together with our dedicated orbital-optimizer, to reduce the cost of quantum simulations.

\begin{figure}
    \centering  
    \includegraphics[width=8cm]{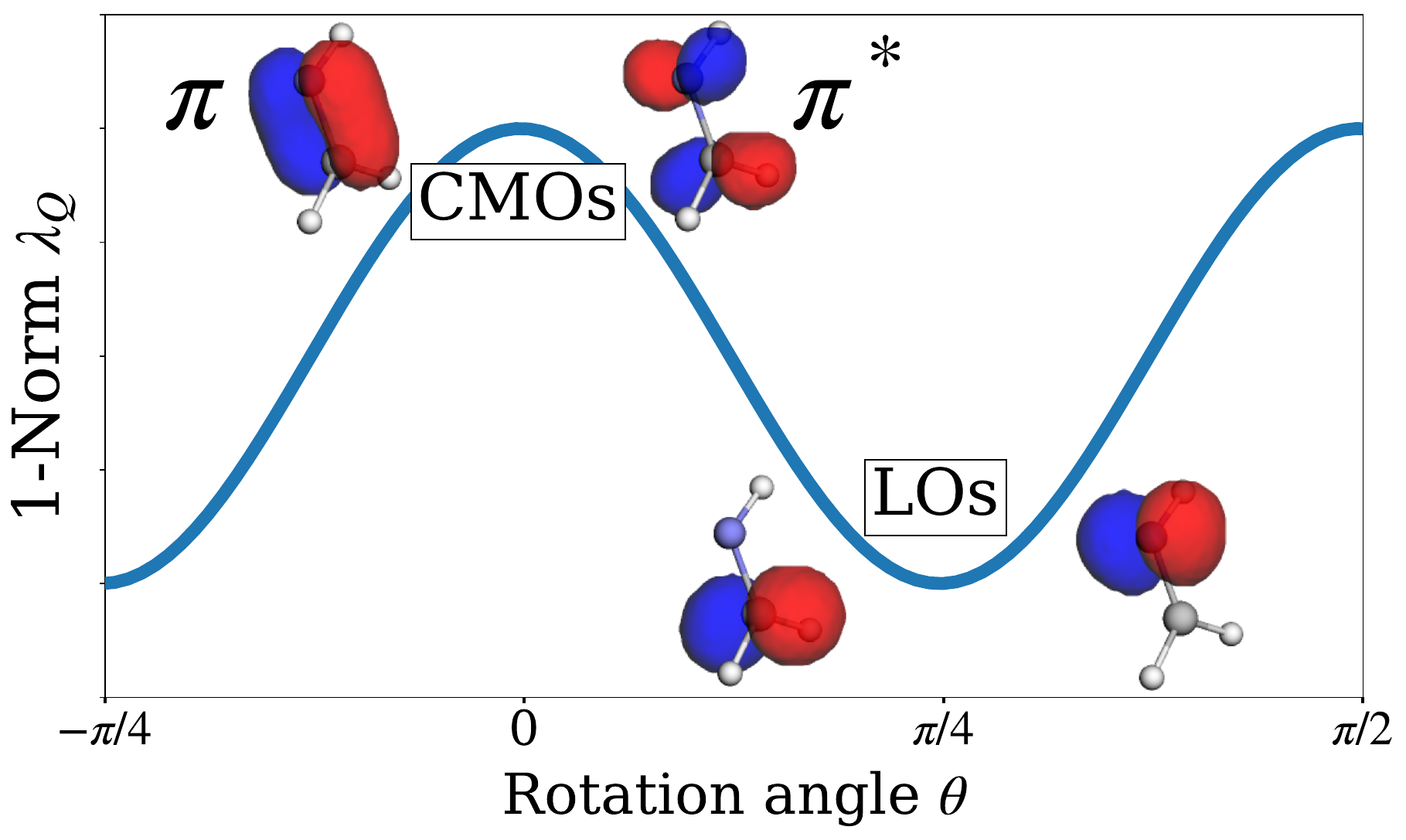}
    \caption{An illustration of how to decrease the 1-Norm on the formaldimine molecule: moving from CMOs to LOs with a Jacobi rotation %$\mathbf{U}(\theta) = \begin{pmatrix}
    % \cos{\theta} & \sin{\theta}\\
    % -\sin{\theta} & \cos{\theta}
    % \end{pmatrix}$
    between occupied $\pi$-bonding and virtual $\pi^*$-antibonding orbitals.}
    \label{fig:jacobi_example}
\end{figure}

For the localization schemes it is usual that the bigger the active space is, the more $\lambda_Q$ can be reduced compared to CMOs.
This can be rationalized as follows.
When the active space increases, it may involve orbitals localized on atoms which have negligible overlap with the other orbitals.
Another possibility is that the additional orbitals may have a positive effect on the localization procedure. As an illustration, see the example shown in Fig.~\ref{fig:jacobi_example} where we consider an occupied $\pi$-bonding orbital mixing with its associated virtual $\pi^*$-antibonding orbital for the formaldimine molecule.
By rotating those two orbitals via a Jacobi rotation, we move from a maximal value of $\lambda_Q$ for $\theta = 0$ (corresponding to the original CMOs)
to a minimal value for $\theta = \pi/4$ (corresponding to the new LOs).
This way, we see that the larger is the active space the more degrees of freedom we have to perform a better orbital localization. For the biggest active space chosen, an active space of 100 electrons in 100 orbitals, $\lambda_Q$ is reduced by an order of magnitude. This illustrates the potential of using widely available and computationally cheap classical pre-optimizations, together with our dedicated orbital-optimizer, to reduce the cost of quantum simulations.

% The 1-norm, and thereby the expected
% cost of the quantum simulation, can be brought down further by using a dedicated optimizer, but it is debatable
% whether the modest further reduction warrants the increased cost of the classical pre-processing.

% In calcuting $\lambda_Q$, the constant term in the Hamiltonian (i.e. $|h_p|$ for which $\hat{P}_p = \mathcal{I}$) is ignored, as we can classically add this term to the energy.

%\makeatletter\onecolumngrid@push\makeatother
\begin{table*}%[ht]
\begin{tabular}{ |p{2cm}||p{0.9cm}p{0.9cm}p{0.9cm}|p{0.9cm}p{0.9cm}p{0.9cm}|p{0.9cm}p{0.9cm}p{0.9cm}|p{0.9cm}p{0.9cm}p{0.9cm}|p{0.9cm}p{0.9cm}p{0.9cm}|p{0.9cm}p{0.9cm}p{0.9cm}|}
 \hline
 \thead{Molecule}     & \thead{$\alpha_{\rm CMO}$} & \thead{$\beta_{\rm CMO}$} & \thead{$R^2_{\rm CMO}$} & \thead{$\alpha_{\rm PM}$} & \thead{$\beta_{\rm CMO}$} & \thead{$R^2_{\rm PM}$} & \thead{$\alpha_{\rm FB}$} & \thead{$\beta_{\rm CMO}$} & \thead{$R^2_{\rm FB}$} & \thead{$\alpha_{\rm ER}$} & \thead{$\beta_{\rm CMO}$} & \thead{$R^2_{\rm ER}$} & \thead{$\alpha_{\rm OO}$} & \thead{$\beta_{\rm OO}$}& \thead{$R^2_{\rm OO}$} \\
 \hline
 FeMoco   &  2.83 &-3.37 & 0.9996 &  2.03 &-1.60 & 0.9477 & 1.94 &-1.26  & 0.9580 & 1.98 & -1.49 & 0.949 & \textbf{1.89} & -1.2 & 0.9688\\
 Ru SI I & 2.63 & -2.84 & 0.9957 & 2.36 &  -3.13  &  0.9465  & \textbf{2.17} & -2.47 &  0.9865 &  2.25 &  -2.78 & 0.9763 & 2.21  &   -2.80 &  0.9927\\
 Ru SI II & 2.83 & -3.71 & 0.9997 &   2.27  & -2.81 & 0.9530  &   \textbf{2.11} & -2.26 &  0.9867 &  2.22 & -2.68 &  0.9770 & 2.20  & -2.8  & 0.9915\\
 Ru TS II-III  & 2.95 & -4.22  &  0.9998 &  2.16  &  -2.29 & 0.8919 &  \textbf{1.98} &  -1.66 & 0.9659 &  2.13 & -2.26 &  0.9234 &  2.12 & -2.44  & 0.9946\\
 Ru SI V & 2.75  & -3.41  & 0.9979 &  2.42 & -3.37 & 0.9063 &  \textbf{2.13} & -2.33 &  0.9798 & 2.33 & -3.09  & 0.9526 & 2.21  &  -2.83 & 0.9926\\
 Ru SI VIII & 2.84 & -3.79  &  0.9989  &  2.22  & -2.54  & 0.8630 &  \textbf{2.12} & -2.24 & 0.9659 &  2.18 & -2.48 & 0.9323 & 2.19   & -2.73  & 0.9901\\
 Ru TS VIII-IX & 2.82 & -3.57  & 0.9995 &  2.23 &  -2.51  &  0.8645 &  \textbf{2.09} & -2.07 &  0.9706 &  2.19  & -2.44 & 0.9039 & 2.19  & -2.66  & 0.9883\\
 Ru SI IX & 2.8 & -3.66  &  0.9987 &  2.42  &  -3.36 &  0.9240 &  \textbf{2.22} & -2.69 & 0.9876  & 2.34 & -3.14 & 0.9663 & 2.23  & -2.91  & 0.9967\\
 Ru SI XVIII & 2.79  &  -3.46 & 0.9986 &  2.31 &  -2.87 & 0.9302  & \textbf{2.12} & -2.16 & 0.9854 & 2.26 & -2.74 & 0.9557 & 2.24  &  -2.87 &  0.9929\\
 \hline
\end{tabular}
\caption{Scaling of $\lambda_Q$ with respect to the active space size $N_{\rm act}$ for different orbital bases, assuming $\log{\lambda_Q}=\alpha \log{N_{\rm act}} +\beta$ with the associated $R^2$ value. The minimal ANO-RCC basis-set is considered.\label{tab:scaling}}
\end{table*}

\section{Conclusions}\label{sec:conclusion}

Numerically we have seen that exploiting the locality of the basis gives rise to a lower variance in the Hamiltonian coefficients, reducing $\lambda_Q$. This results in a significant reduction of the absolute values of the integrals. The off-diagonal elements of the two-electron tensor play the biggest role here.

To find an even better basis-rotation that minimizes $\lambda_Q$, one can use a ``brute-force" orbital optimizer as we have done in this work. This method could be realized in practice only because we have a efficient way of computing $\lambda_Q$ directly in terms of the molecular integrals [Eq.~\eqref{eq:1-norm_Q_pqrs}], avoiding doing a fermion-to-qubit mapping explicitly. Expressing it as $\lambda_Q = \lambda_T + \lambda^{\prime}_V$, this improves a bit over the equation of $\lambda_V$ in Ref.~\onlinecite{lee2020even}, holding into account the representation of the Hamiltonian in terms of unique Pauli strings. The direct orbital optimizer can be used in quite large active spaces, reliably converging for active spaces with up to $N_{\rm act} \approx 80$ orbitals, sometimes more. For the largest active spaces considered, it converges very slowly due to the need of 4-index transformations at every step which are costly because of the large dimensionality of the system, combined with the non-continuous and complicated landscape of $\lambda_Q$. This gives rise to the need to either set a high threshold of convergence or cut off the optimizer at some point where the user is satisfied with the result. While this is problematic, there is no a priori reason not to try this optimizer on the (potentially large) active spaces of the molecule one is considering. This can be a point of further study. 
% \emiel{maybe need to rephrase this}

We benchmarked a range of various molecules and active spaces for which we showed to achieve a significant reduction of $\lambda_Q$. Apart from this benchmark, we investigated the scaling of $\lambda_Q$ with the size of the system $N$. There are multiple paths one can take here to increase $N$. A popular way to do this in the literature is considering either a hydrogen chain or hydrogen ring, and increase the amount of atoms. We benchmarked the scaling of $\lambda_Q$ for a hydrogen chain and a linear alkane chain by increasing the number of atoms as well, where our results show a scaling of $\mathcal{O}(N^{2.3})$ and $\mathcal{O}(N^{2.2})$, respectively. Here we saw localized orbitals can give approximately a factor of $N$ lower scaling of $\mathcal{O}(N^{1.3})$ and $\mathcal{O}(N^{1.4})$, respectively. As this gives one limited information how big $\lambda_Q$ will be in real eventual applications of quantum algorithms, we decided to investigate the scaling on relevant highly correlated molecules such as FeMoco (important in nitrogen fixation) and Ruthenium metal complexes (important in carbon dioxide capture). Even though we used a minimal basis to make the scaling calculations feasible on these molecules, there is no reason to believe $\lambda_Q$ will scale differently with respect to the active space size on a bigger basis-set. Here we showed that the scaling is a factor of $N$ larger, and with localized orbitals can be brought down to $\lambda_Q \approx \mathcal{O}(N^2)$, depending on the molecule in consideration. Our dedicated orbital-optimizer was able to consistently give an even further improvement of 10-25\% on these molecules. As a concrete example, consider the largest active space considered of FeMoco: this factor of $N$ difference would result in a reduction by two orders of magnitude in the amount of measurements needed for tomography [Eq.~\eqref{eq:vqe_measure}], if one wants chemical accuracy.
This indicates that the simple and efficient classical pre-processing by widely available localization techniques, together with our dedicated optimizer, will help to make simulations with large active spaces feasible.
%\newpage
%------------------------------------------------
%\phantomsection
\section*{Acknowledgments} % The \section*{} command stops section numbering
We sincerely thank Susi Lethola for his helpful comments on localization schemes.
SY and BS acknowledge support from the Netherlands Organization for Scientific Research (NWO/OCW). EK acknowledges support from Shell Global Solutions BV.

% \luuk{Funders: Shell-Emiel, NWO-Saad. Check previous papers,should be able to copy-paste from there}

%----------------------------------------------------------------------------------------

%\addcontentsline{toc}{section}{Acknowledgments} % Adds this section to the table of contents
%\clearpage
% =======================================================
% =======================================================
%                   A P P E N D I X
% =======================================================
% =======================================================
\appendix

\section{Frozen core Hamiltonian}\label{appendix:FrozenCore}

Applying the frozen core approximation to the electronic structure Hamiltonian consists in assuming the existence of a set of frozen orbitals (always occupied), another set of active orbitals (belonging to an active space), and a set of virtual orbitals (always unoccupied). 
Based on this partitioning, every Slater determinant $|\Phi\rangle$ used to describe properties of the system take the form 
\begin{equation}
    |\Phi\rangle = | \Phi_\text{frozen}\Phi_\text{active} \rangle,
    \label{eq:state}
\end{equation}
where the left contribution $\Phi_{\rm frozen}$ is the part of the determinant encoding the frozen orbitals of the system (always occupied) whereas $\Phi_{\rm active}$ is a part encoding the occupancy of the remaining electrons in the active orbitals of the system. 
In this context, if one considers that every correlated electronic wavefunction is always expanded with Slater determinants following Eq.~(\ref{eq:state}), one can demonstrate by projections that the system Hamiltonian takes an effective form 
\begin{eqnarray}\label{eq:Ham_elec}
\bra{\Phi} {\mathcal{\hat{H}}} \ket{\Phi} \equiv \bra{\Phi_{\rm active}} {\mathcal{\hat{H}}}^{\rm FC} \ket{\Phi_{\rm active}},
\end{eqnarray}
with ${\mathcal{\hat{H}}}^{\rm FC}$ the so-called ``frozen core Hamiltonian'' defined as follows,
\begin{eqnarray}
\label{eq:Ham_FC}
{\mathcal{\hat{H}}}^{\rm FC} = {\mathcal{\hat{H}}}_\text{active} + E_\text{frozen}^\text{MF} + \mathcal{\hat{V}}.
\end{eqnarray}
Here, ${\mathcal{\hat{H}}}_\text{\rm active}$ is the Hamiltonian encoding the one- and two- body terms only acting in the active space,
\begin{equation}
\label{eq:AS_HAM}
{\mathcal{\hat{H}}}_\text{active} = \sum_{tu}^\text{active} h_{tu} \hat{E}_{tu} + \sum_{tuvw}^\text{active} g_{tuvw} \hat{e}_{tuvw},
\end{equation}
where $t,u,v,w$ denote active space orbitals.
The second term $E_{\rm frozen}^{\rm MF}$ is a scalar representing the mean-field-like energy obtained from the frozen orbitals,
\begin{equation}
    E_\text{frozen}^\text{MF} = 2\sum_i^\text{frozen} h_{ii} + \sum_{ij}^\text{frozen} (2g_{iijj}- g_{ijji}),
    \label{eq:shift}
\end{equation}
and the third term
\begin{equation}
\label{eq:emb}
\mathcal{\hat{V}} = \sum_{tu}^\text{active} \mathcal{V}_{tu} \hat{E}_{tu} \text{, with } \mathcal{V}_{tu} =  \sum_i^\text{frozen} (2g_{tuii}- g_{tiiu} )
\end{equation}
represents an effective one body potential which encodes the interaction of the frozen electrons with the active space electrons. To summarize, the main effect of the frozen core approximation [Eq.~(\ref{eq:Ham_FC})] is first to introduce an energetic shift [Eq.~(\ref{eq:shift})], and second to augment the one body term of the Hamiltonian operator [Eq.~(\ref{eq:AS_HAM})] (that only lives in the active space) with an additional effective one body operator [Eq.~(\ref{eq:emb})].

\section{Electronic integrals transformation}
\label{appendix:HamTransfo}

Preparing the electronic Hamiltonian in a given orthogonal orbital basis $\lbrace \phi_p(\mathbf{r}) \rbrace_{p=1}^N$ is a crucial step for realizing concrete quantum computing applications. In practice, such a process can be realized classically via electronic integral transformations. For this, we assume that the orthogonal orbitals can be expressed as a linear combination of AOs such that
\begin{equation}
    \phi_p(\mathbf{r}) = \sum_\mu  \chi_\mu(\mathbf{r}) C_{\mu p},
\end{equation}
where we assume that the functions $\chi_\mu(\mathbf{r})$ as well as the coefficient matrix $\mathbf{C}$ are real-valued, as is usually done in non-relativistic quantum chemistry.
Based on our knowledge of this matrix $\mathbf{C}$, one can transform the one- and two-electron integrals from the AO basis to the orthogonal orbital basis. To do so, one starts from the one- and two-electron integrals in the AO basis, respectively $h_{\mu\nu}$ and $g_{\mu\nu\gamma\delta}$, and implements the following two- and four-indexes transformations:
\begin{equation}
 h_{pq} =  \sum_{\mu\nu} h_{\mu\nu} C_{\mu p}C_{\nu q},\label{eq:one_elec_new}
\end{equation}
and 
\begin{equation}
 g_{pqrs} =  \sum_{\mu\nu\gamma\delta} g_{\mu\nu\gamma\delta} C_{\mu p}C_{\nu q}C_{\gamma r}C_{\delta s}.\label{eq:two_elec_new}
\end{equation}
From a computational point of view, the numerical cost of processing electronic integrals is essentially governed by the four-indexes transformation. 
This transformation is known to scale as $\mathcal{O}(N^5)$ (see Ref.~\onlinecite{yoshimine1973construction}), when not employing approximations such as density fitting (see Ref.~\onlinecite{feyereisen1993use}).

\section{Explicit form of $\lambda_Q$ in terms of molecular integrals}\label{appendix:lambda_Q}

To derive a formula for $\lambda_Q$ in terms of the molecular integrals, one needs to keep track of what happens to the coefficients in the Hamiltonian (Eq.~\eqref{eq:Ham_elec}) after a fermion-to-qubit transformation of the fermionic operators.
This is non-trivial, since e.g. the Jordan-Wigner transformation transforms a single arbitrary product of fermionic operators $\hat{a}_p^{\dagger} \hat{a}_r^{\dagger} \hat{a}_s^{} \hat{a}_q^{}$ into 16 different Pauli strings, because of the products of $\frac{\hat{X} - i \hat{Z}}{2}$ and $\frac{\hat{X} + i \hat{Z}}{2}$.
The Hamiltonian in the form of \eqref{eq:Ham_elec} is actually not best suited for this derivation. 
Instead, it is helpful to express the Hamiltonian in terms of Majorana operators
that have the convenient property of being Hermitian operators that square to identity:
\begin{align}\label{eq:major}
\hat{\gamma}_{p\sigma,0} = \hat{a}_{p\sigma}^{} + \hat{a}_{p\sigma} ^{\dagger},  \quad
\hat{\gamma}_{p\sigma,1} = -i \left( \hat{a}_{p\sigma}^{} - \hat{a}_{p\sigma}^{\dagger}\right),\\
\left\{\hat{\gamma}_i,\hat{\gamma}_j\right\} = 2\delta_{ij}\mathcal{I}, \quad \hat{\gamma}_i^\dagger = \hat{\gamma}_i, \quad \hat{\gamma}_i^2 = \mathcal{I}.
\end{align}
To find a representation of the Hamiltonian in terms of Majorana operators, one could directly replace $\hat{a}_p^\dagger = \frac{\hat{\gamma}_{p,0} - i\hat{\gamma}_{p,1}}{2}$ and $\hat{a}_q^{} = \frac{\hat{\gamma}_{q,0} + i \hat{\gamma}_{q,1}}{2}$ in the Hamiltonian, or follow the procedure in Ref.~\onlinecite{von2020quantum}.
We chose the latter approach, detailed below.

It is well known that the electron repulsion integral tensor, when written as a $N^2\times N^2$ matrix $g_{(pq),(rs)}$ with the composite indices $pq$ and $rs$, is positive semi-definite. This makes it possible to define a Cholesky decomposition 
 $\mathbf{g} = \mathbf{L}\mathbf{L}^\dagger$. We write:
\begin{align}
    g_{pqrs} = \sum_\ell \sum_{pqrs} L_{pq}^\ell L_{rs}^\ell.
\end{align}
with $\mathbf{L}$ a lower triangular matrix.
Since $g_{pqrs}$ is symmetric in $p,q$ and in $r,s$, for a given $\ell$, $L_{pq}^\ell$ is an $N \times N$ symmetric matrix (also called a Cholesky vector in the Cholesky decomposition).
This leads to the following expression for the Hamiltonian:
\begin{align}\label{eq:Ham_CD}
\begin{split}
    \mathcal{\hat{H}} = \sum_{pq}&\sum_\sigma h_{pq}\hat{a}_{p\sigma}^\dagger\hat{a}_{q\sigma}^{} + \dfrac{1}{2} \sum_{pqrs}\sum_{\sigma \tau} g_{pqrs}\hat{a}_{p\sigma}^\dagger\hat{a}_{r\tau}^\dagger\hat{a}_{s\tau}^{}\hat{a}_{q\sigma}^{} \\
    = \sum_{pq \sigma}& \left[h_{pq} - \frac{1}{2}\sum_{r}g_{prrq}\right]\hat{a}_{p\sigma}^\dagger\hat{a}_{q\sigma}^{}\\
    &+ \dfrac{1}{2} \sum_\ell \sum_{pqrs} \sum_{\sigma \tau} L_{pq}^\ell L_{rs}^\ell\hat{a}_{p\sigma}^\dagger\hat{a}_{q\sigma}^{}\hat{a}_{r\tau}^\dagger\hat{a}_{s\tau}^{} \\
    = \sum_{pq \sigma}& h_{pq}^\prime\hat{a}_{p\sigma}^\dagger\hat{a}_{q\sigma}^{}
    + \dfrac{1}{2} \sum_\ell\left(\sum_{pq\sigma} L_{pq}^\ell\hat{a}_{p\sigma}^\dagger\hat{a}_{q\sigma}^{}\right)^2
\end{split}
\end{align}
where $h_{pq}^\prime = h_{pq} - \frac{1}{2}\sum_{r}g_{prrq}$.
Using the relation
\begin{align}
   \hat{a}_{p\sigma}^{\dagger}\hat{a}_{q\sigma}+\hat{a}_{q\sigma}^{\dagger}\hat{a}_{p \sigma}=\left\{\begin{array}{ll}
\mathcal{I}+i\left(\hat{\gamma}_{p\sigma, 0} \hat{\gamma}_{p\sigma, 1}\right), & p=q \\
\frac{i}{2}\left(\hat{\gamma}_{p\sigma, 0} \hat{\gamma}_{q\sigma, 1}+\hat{\gamma}_{q\sigma, 0} \hat{\gamma}_{p\sigma, 1}\right), & p \neq q
\end{array}\right.
\end{align}
one can show that
\begin{align}\label{eq:L_majorana}
    \sum_{pq\sigma} M_{pq}\hat{a}_{p\sigma}^{\dagger} \hat{a}_{q\sigma}^{} = \sum_{p}M_{pp} \mathcal{I} + \frac{i}{2} \sum_{pq\sigma}M_{pq} \hat{\gamma}_{p\sigma,0}\hat{\gamma}_{q\sigma,1},
\end{align}
where $M_{pq}$ can be replaced by any symmetric matrix (like $L_{pq}^\ell$ or $h_{pq}^\prime$). 
Employing Eq.~\eqref{eq:L_majorana} in Eq.~\eqref{eq:Ham_CD},
leads to the expression of the Hamiltonian in terms of Majorana operators:
\begin{align}
\begin{split}
    \mathcal{\hat{H}} =& \sum_{p} h_{pp}^{\prime}\mathcal{I} + \dfrac{i}{2}\sum_{pq\sigma} h_{pq}^{\prime} \hat{\gamma}_{p\sigma,0}\hat{\gamma}_{q\sigma,1}\\
    &+ \frac{1}{2}\sum_{\ell} \left(\sum_{p}L_{pp}^\ell \mathcal{I} + \frac{i}{2} \sum_{pq\sigma}L_{pq}^\ell \hat{\gamma}_{p\sigma,0}\hat{\gamma}_{q\sigma,1}\right)^2.
\end{split}
\end{align}
After working out the square, and substituting back $\sum_\ell \sum_{pqrs} L_{pq}^\ell L_{rs}^\ell = g_{pqrs}$, we obtain
\begin{align}\label{eq:Ham_maj1}
\begin{split}
    \mathcal{\hat{H}} =& \left( \sum_p h_{pp} + \frac{1}{2}\sum_{pr} g_{pprr} - \frac{1}{2}\sum_{pr}g_{prrp}\right)\mathcal{I} \\
    &+\frac{i}{2}\sum_{pq\sigma}\left(h_{pq} + \sum_r g_{pqrr} - \frac{1}{2} \sum_r g_{prrq}\right)\hat{\gamma}_{p\sigma,0}\hat{\gamma}_{q\sigma,1}\\
    &-\frac{1}{8}\sum_{pqrs\sigma\tau} g_{pqrs} \hat{\gamma}_{p\sigma,0}\hat{\gamma}_{q\sigma,1}\hat{\gamma}_{r\tau,0}\hat{\gamma}_{s\tau,1}.
\end{split}
\end{align}

We now want to determine the value of $\lambda_Q$ after a fermion-to-qubit mapping of this Hamiltonian in which all products of Majorana operators should be distinct. We use a fermion-to-qubit mapping such that single Majorana operators are mapped to single unique Pauli strings, such as the Jordan-Wigner transformation \cite{jordanwigner}:
\begin{align}
    \hat{\gamma}_{i,0} \rightarrow \hat{X}_{i}  \hat{Z}_{i-1}\dots \hat{Z}_0  \text{, and } \hat{\gamma}_{i,1} \rightarrow \hat{Z}_{i}  \hat{Z}_{i-1}\dots \hat{Z}_0,
\end{align}
where we used the composite index $i = p\sigma$. Note this is not only true for the Jordan-Wigner transformation, but also for for example the Bravyi-Kitaev transformation \cite{havlicek2017operator}. The only thing left to do is to make sure that the sum over 4 indices used to evaluate $\lambda_Q$ is indeed truly quartic. Looking at Eq. (\ref{eq:Ham_maj1}), this is not yet the case. If $p\sigma = r\tau$ or $q\sigma =s\tau$ the corresponding Majorana operators will square to identity and will reduce to quadratic and identity terms. Below we employ the permutational symmetry of the real-valued integrals $g_{pqrs}$  to obtain the most compact representation.

We first reorder the Majorana's and distinguish between the same and opposite spin cases:
\begin{align}
    &-\frac{1}{8}\sum_{pqrs\sigma\tau} g_{pqrs} \hat{\gamma}_{p\sigma,0}\hat{\gamma}_{q\sigma,1}\hat{\gamma}_{r\tau,0}\hat{\gamma}_{s\tau,1}\nonumber\\
    =&\frac{1}{8} \sum_{pqrs}\sum_{\sigma\tau}g_{pqrs}\hat{\gamma}_{p\sigma,0}\hat{\gamma}_{r\tau,0}\hat{\gamma}_{q\sigma,1}\hat{\gamma}_{s\tau,1}\nonumber\\
    =&\frac{1}{8} \sum_{pqrs}\sum_{\sigma}g_{pqrs}\hat{\gamma}_{p\sigma,0}\hat{\gamma}_{r\sigma,0}\hat{\gamma}_{q\sigma,1}\hat{\gamma}_{s\sigma,1}\nonumber\\
    +&\frac{1}{8} \sum_{pqrs}\sum_{\sigma \neq \tau}g_{pqrs}\hat{\gamma}_{p\sigma,0}\hat{\gamma}_{r\tau,0}\hat{\gamma}_{q\sigma,1}\hat{\gamma}_{s\tau,1}
    \label{eq:twobody_notq}
\end{align}
For the opposite spin case all operator products are unique and quartic, so no further work is needed. For the same spin case it is useful to identify cases in which two or more indices of the same type of Majorana operators are equal:

\begin{align}
    &\frac{1}{8} \sum_{pqrs}\sum_{\sigma}g_{pqrs}\hat{\gamma}_{p\sigma,0}\hat{\gamma}_{r\sigma,0}\hat{\gamma}_{q\sigma,1}\hat{\gamma}_{s\sigma,1}\nonumber\\
    =&\frac{1}{8} \sum_{pq}\sum_{\sigma}g_{pqpq}\hat{\gamma}_{p\sigma,0}\hat{\gamma}_{p\sigma,0}\hat{\gamma}_{q\sigma,1}\hat{\gamma}_{q\sigma,1}\nonumber\\
    +&\frac{1}{8} \sum_{p\neq r,q}\sum_{\sigma}g_{pqrq}\hat{\gamma}_{p\sigma,0}\hat{\gamma}_{r\sigma,0}\hat{\gamma}_{q\sigma,1}\hat{\gamma}_{q\sigma,1}\nonumber\\
     +&\frac{1}{8} \sum_{q\neq s,p}\sum_{\sigma}g_{pqps}\hat{\gamma}_{p\sigma,0}\hat{\gamma}_{p\sigma,0}\hat{\gamma}_{q\sigma,1}\hat{\gamma}_{s\sigma,1}\nonumber\\
     +&\frac{1}{8} \sum_{p\neq r,q\neq s}\sum_{\sigma}g_{pqrs}\hat{\gamma}_{p\sigma,0}\hat{\gamma}_{r\sigma,0}\hat{\gamma}_{q\sigma,1}\hat{\gamma}_{s\sigma,1}\nonumber\\
     =&\frac{1}{4} \sum_{pq}g_{pqpq}\mathcal{I}\nonumber\\
     +&\frac{1}{8} \sum_{p\neq r,q\neq s}\sum_{\sigma}g_{pqrs}\hat{\gamma}_{p\sigma,0}\hat{\gamma}_{r\sigma,0}\hat{\gamma}_{q\sigma,1}\hat{\gamma}_{s\sigma,1},
    \label{eq:twobody_samespin}
\end{align}

where we applied the identity relation $\hat{\gamma}_i^2=\mathcal{I}$ and for the second term used that integrals are symmetric in exchanging $p$ and $r$ while the product of Majorana operators is antisymmetric under this exchange. This nullifies this and the third term. The first term can be absorbed in the scalar term. For the truly quartic term we may
make further  use of permutational symmetries to get the most compact form:

\begin{align}
    &\frac{1}{8} \sum_{p\neq r,q\neq s}\sum_{\sigma}g_{pqrs}\hat{\gamma}_{p\sigma,0}\hat{\gamma}_{r\sigma,0}\hat{\gamma}_{q\sigma,1}\hat{\gamma}_{s\sigma,1}\nonumber\\
    =&\frac{1}{8} \sum_{p>r, s>q}\sum_{\sigma}(\nonumber\\
    & g_{pqrs}\hat{\gamma}_{p\sigma,0}\hat{\gamma}_{r\sigma,0}\hat{\gamma}_{q\sigma,1}\hat{\gamma}_{s\sigma,1}\nonumber\\
    +&g_{rqps}\hat{\gamma}_{r\sigma,0}\hat{\gamma}_{p\sigma,0}\hat{\gamma}_{q\sigma,1}\hat{\gamma}_{s\sigma,1}\nonumber\\
    +&g_{psrq}\hat{\gamma}_{p\sigma,0}\hat{\gamma}_{r\sigma,0}\hat{\gamma}_{s\sigma,1}\hat{\gamma}_{q\sigma,1}\nonumber\\
    +&g_{rspq}\hat{\gamma}_{r\sigma,0}\hat{\gamma}_{p\sigma,0}\hat{\gamma}_{s\sigma,1}\hat{\gamma}_{q\sigma,1})\nonumber\\
        =&\frac{1}{4} \sum_{p>r, s>q}\sum_{\sigma}(g_{pqrs}-g_{psrq})\hat{\gamma}_{p\sigma,0}\hat{\gamma}_{r\sigma,0}\hat{\gamma}_{q\sigma,1}\hat{\gamma}_{s\sigma,1}
    \label{eq:twobody_finalstep}
\end{align}

Combining all terms we end up with the expression:

\begin{align}
\begin{split}
    \mathcal{\hat{H}} =& \left(\sum_p^N h_{pp} + \frac{1}{2}\sum_{pr}^N g_{pprr} - \frac{1}{4}\sum_{pr}^N g_{prrp} \right)\mathcal{I} \\
    &+ \frac{i}{2}\sum_{p q \sigma}^N\left(h_{pq} + \sum_r^N g_{pqrr} - \frac{1}{2} \sum_r^N  g_{prrq}\right) \hat{\gamma}_{p\sigma,0}\hat{\gamma}_{q\sigma,1}\\
    &+ \frac{1}{4}\sum_{p>r, s>q}^N \sum_{\sigma} \left(g_{pqrs} - g_{psrq}\right)\hat{\gamma}_{p\sigma,0}\hat{\gamma}_{r\sigma,0}\hat{\gamma}_{q\sigma,1}\hat{\gamma}_{s\sigma,1}\\
    &+ \frac{1}{8} \sum_{pqrs}^N\sum_{\sigma\neq\tau}g_{pqrs}\hat{\gamma}_{p\sigma,0}\hat{\gamma}_{r\tau,0}\hat{\gamma}_{q\sigma,1}\hat{\gamma}_{s\tau,1}.
\end{split}
\label{eq:Ham_maj2}
\end{align}

We then take the sum of absolute values of the coefficients and perform the sums over spin in the quartic term explicitly (amounting to a factor of 2), to get the following form of $\lambda_Q$:
\begin{eqnarray}\label{eq:appendix_lambdaQ}
    \lambda_Q &=& \lambda_C + \lambda_T + \lambda^{\prime}_V,
\end{eqnarray}
where $\lambda_C$ corresponds to the constant term in Eq.~\eqref{eq:Ham_maj2}, $\lambda_T$ to the quadratic and $\lambda_V$ to the quartic term in Majorana operators. They have the form:
\begin{eqnarray}
\lambda_C &=&\left|\sum_p^N h_{pp} + \frac{1}{2}\sum_{pr}^N g_{pprr} - \frac{1}{4}\sum_{pr}^N g_{prrp}\right| \label{eq:appendix_lambda_C},\\
\lambda_T &=& \sum_{p q}^N\left|h_{pq} + \sum_r^N g_{pqrr} - \frac{1}{2} \sum_r^N g_{prrq}\right|, \label{eq:appendix_lambda_T}\\
\lambda^{\prime}_V &=& \frac{1}{2}\sum_{p>r, s>q}^N\left|g_{pqrs} - g_{psrq}\right|+ \frac{1}{4} \sum_{pqrs}^N|g_{pqrs}| \label{eq:appendix_lambda_V}.
\end{eqnarray}

This $\lambda^{\prime}_V$ is slightly different to the $\lambda_V$ of Ref.~\onlinecite{lee2020even} in calculating the $\lambda$ for the "sparse" algorithm of Berry et al. \cite{berry2019qubitization}, which should actually be the same as $\lambda_Q$. This is due that we took into account the swapping of majorana operators in Eq.~\eqref{eq:twobody_finalstep}. In Ref.~\onlinecite{lee2020even}, this corresponds to the term $\hat{V}^{\prime}=\frac{1}{8} \sum_{\alpha, \beta \in\{\uparrow, \downarrow\}} \sum_{p, q, r, s} V_{p q r s} \hat{Q}_{p q \alpha} \hat{Q}_{r s \beta}$. It is hard to see here that the product $\hat{Q}_{p q \alpha} \hat{Q}_{r s \alpha}$ is anti-symmetric in swapping $p,r$ and $q, s$, but it becomes clear when one realizes that
$\hat{Q}_{p q \alpha} = i\hat{\gamma}_{p\sigma,0}\hat{\gamma}_{q\sigma,1}$, indicating the usefulness of majorana operators. As the absolute values give that:
\begin{align}
    |g_{pqrs} - g_{psrq}| \leq |g_{pqrs}| + |g_{psrq}| \,\, \forall \,\, p,q,r,s
\end{align}
such that $\lambda^{\prime}_V < \lambda_V$  (except when $g_{pqrs}$ and $g_{psrq}$ always have opposite sign, in which case they would be equal).

%----------------------------------------------------------------------------------------
%	REFERENCE LIST
%----------------------------------------------------------------------------------------
\phantomsection
\bibliography{biblio}

\end{document}